\documentclass[12pt]{article}
	\usepackage[a4paper, lmargin=1.25in,rmargin=1.25in,tmargin=1.25in,bmargin=1.25in]{geometry}

	\usepackage{amsmath}
	\usepackage{amssymb}
	\usepackage[mathscr]{eucal}
	\usepackage{mathtools}
	\usepackage{graphicx}
	\usepackage{pgfplots}
	\usepgfplotslibrary{groupplots}
	\usepackage{dsfont}
	\usepackage{bm}
	\usepackage[T1]{fontenc}

	\usepackage{setspace}
	\usepackage[font=footnotesize,margin=0in]{caption}
	\usepackage{enumitem}
	\usepackage{hyperref}
	\hypersetup{
	    colorlinks=true,
	    citecolor=blue,
	    linkcolor=blue,
	    urlcolor=blue,
	}
	\usepackage[capitalise]{cleveref}
	\usepackage{amsthm}
	\newtheorem{proposition}{Proposition}
	\newtheorem{theorem}{Theorem}
	\newtheorem{corollary}{Corollary}
	\newtheorem{lemma}{Lemma}
	\newtheorem{remark}{Remark}
	\theoremstyle{definition}
	\newtheorem{definition}{Definition}

	\Crefname{claim}{Claim}{Claims}

	\newcommand{\setword}[2]{%
	  \phantomsection
	  #1\def\@currentlabel{\unexpanded{#1}}\label{#2}%
	}

	\usepackage[backend=biber,authordate,useprefix=true]{biblatex-chicago}
	\usepackage[page]{appendix}
	\crefname{app}{Appendix}{Appendices}
	\crefname{supp}{Online Appendix}{Online Appendices}

	\let\originalleft\left
	\let\originalright\right
	\renewcommand{\left}{\mathopen{}\mathclose\bgroup\originalleft}
	\renewcommand{\right}{\aftergroup\egroup\originalright}

	\usepackage{scalerel,stackengine}

	\makeatletter
	\newcommand{\pushright}[1]{\ifmeasuring@#1\else\omit\hfill$\displaystyle#1$\fi\ignorespaces}
	\newcommand{\pushleft}[1]{\ifmeasuring@#1\else\omit$\displaystyle#1$\hfill\fi\ignorespaces}
	\makeatother

	\DeclareMathOperator*{\argmax}{arg\,max}

	\title{
		Incentives for Collective Innovation%
		\thanks{
			I thank my advisors
			P\'{e}ter Es\H{o} and
			John Quah, 
			as well as 
			Martin Cripps, 
			Francesc Dilm\'e, 
			Matteo Escud\'e, 
			George Georgiadis, 
			Claudia Herresthal, 
			Mathijs Janssen,
			Sam Jindani, 
			Philippe Jehiel,
			Meg Meyer, 
			Sven Rady, 
			Ludvig Sinander,  
			and the seminar participants at 
			Aalto, 
			Bonn, 
			Essex,
			Oxford, 
			PSE, 
			Venice and 
			Warwick 
			for helpful comments.
			I gratefully acknowledge support by the German Research Foundation (DFG) through CRC TR 224 (Project B02).
		}
	}
	\author{
		Gregorio Curello%
		\footnote{
			University of Mannheim.
			E-mail: \href{mailto:gregorio.curello@uni-mannheim.de}{gregorio.curello@uni-mannheim.de}.
		}
	}
	
\begin{document}

	\pgfmathsetmacro{\l}{10}
	\pgfmathsetmacro{\r}{0.01}
	\pgfmathsetmacro{\rmed}{0.0334}
	\pgfmathsetmacro{\rlow}{0.06666}
	\pgfmathsetmacro{\e}{0.01}
	\pgfmathsetmacro{\z}{5}
	\pgfmathsetmacro{\n}{5}
	\pgfmathsetmacro{\lp}{0.1}
	\pgfmathsetmacro{\ep}{1}
	\pgfmathsetmacro{\re}{0.05}

	\pgfmathsetmacro{\xstar}{\l*\n*(\e+\r*(\z-\e))}
	\pgfmathsetmacro{\xstarmax}{4}
	\pgfmathsetmacro{\xf}{\l*(\e+\r*(\z-\e))}
	\pgfmathsetmacro{\yf}{0.190904}
	\pgfmathsetmacro{\yfmed}{0.161138066802}
	\pgfmathsetmacro{\yflow}{0.12614770059}
	\pgfmathsetmacro{\xfp}{\lp*(\ep+\z)}
	\pgfmathsetmacro{\yfp}{0.186813}
	\pgfmathsetmacro{\xfhatp}{0.494639}
	\pgfmathsetmacro{\ydp}{\lp*\z/(1 + \lp*(\n - 1))}

	\pgfmathsetmacro{\xe}{\l*\re*(\z-\e)}
	\pgfmathsetmacro{\ye}{(\e*(1+\l*\n)+\xe)/(1+\l*\re*(\n-1))}

	\pgfmathsetmacro{\xfmax}{0.8}

	\maketitle

	\begin{abstract}
		Agents exert hidden effort to produce randomly-sized innovations in a technology they share.
		Flow payoffs grow as the technology develops, but so does the marginal cost of effort. 
		I characterise the unique symmetric MPE with the quality of the technology as the state variable.
		In this equilibrium, continuation payoffs may fall after an innovation.
		I show that this occurs (with positive probability) if the number of agents is sufficiently large.
		Allowing agents to discard innovations induces higher effort at all states in the symmetric MPE (which remains unique).
		Ex-ante payoffs are higher as well and, under natural conditions, they exceed those of all equilibria without disposal.

	  %
	\end{abstract}

	\section{Introduction}
	Innovation often has the features of a collective-action problem, since innovators gradually improve a technology that they share, while simultaneously using it. 
	For example, firms collaborate to refine their products,%
	\footnote{
		Knowledge exchange among firms was key to the development of biotechnologies (\textcite{powell1996}), semiconductors (\textcite{chesbrough2003,lim2009}), and in the steel industry (\textcite{vonHippel1987}).
	}  
	non-profit organisations draw from a pool of shared knowledge to improve their social programs,
		%
		%
	and workers in a team learn from each other how better to perform their tasks.
		%
		%

	However, innovation differs from traditional models of public-good provision in at least two respects.
	First, both the arrival time and the magnitude of innovations are typically uncertain. 
	Second, innovators face a resource trade-off between using and improving the shared technology.
	For example, firms that collaborate towards a technological improvement must allocate funds between its pursuit and private activities that make use of the technology being developed.
	This trade-off is typically due to financial constraints, a known barrier to innovation for firms (\textcite{hottenrott2012}).
	Similarly, non-profits split resources between developing new ideas and managing existing programmes,
		%
		%
	while employees allocate their time and effort between creative and routine behaviour (\textcite{ford1996}).
	%
	%
	Thus, as the shared technology develops and using it becomes more profitable, the opportunity cost of improving it rises.%
	\footnote{
		In the example involving firms, `using' the technology may also include running private R\&D which does not benefit the other firms.
		\textcite{meyer2003} highlights this incentive effect: `With the establishment of a profitable industry, technological uncertainty is reduced and the collective invention process evaporates. Surviving firms run private R\&D.'
		Similarly, \textcite{powell2010} note that `as technological uncertainty recedes, firms develop private R\&D and focus on their own specific applications. Reliance on collective invention accordingly wanes.'
	}

	In this paper, I analyse a game of public-good provision with these two features.
	Long-lived identical agents exert effort to produce lumpy, randomly-sized increments in the stock of a public good (the arrival rate of increments being proportional to effort).
	Effort is hidden but the stock is observable.
	As the stock grows, flow payoffs rise at a decreasing rate, and the marginal cost of effort rises as well.
	In my interpretation of the model, the stock corresponds to the quality of the technology that the agents share, 
	increments in the stock are innovations, 
	and effort measures the quantity of resources invested in the improvement of the technology, the rest being devoted to its exploitation.

	I derive the welfare benchmark (\Cref{proposition:benchmark}) and characterise the unique symmetric Markov perfect equilibrium with the stock as the state variable (\Cref{theorem:equilibrium}).
	The game admits no other strongly symmetric equilibrium (\Cref{proposition:sse}).
	Moreover, no equilibrium yields long-run gains over the symmetric Markov perfect one, since effort stops in all equilibria precisely when the stock reaches the stopping threshold of the single-agent problem (\Cref{proposition:ppe}).

	In the symmetric equilibrium, innovations affect continuation payoffs through two opposing channels. 
	On one hand, they increase flow payoffs. On the other, they lead agents to exert lower effort, thereby delaying subsequent improvements in flow payoffs. 
	Agents neglect the social cost of this delay when reducing effort, and in some cases the negative effect dominates, causing continuation payoffs to fall after the innovation.%
	\footnote{In the welfare benchmark, innovations always increase continuation payoffs.} 
	Such harmful innovations are produced (with positive probability) in sufficiently large teams (\Cref{proposition:detrimental_n}).
	To build intuition for this result, note that since effort stops at the single-agent threshold, a large enough innovation must increase continuation payoffs. As a consequence, if such innovations are feasible and innovations are produced sufficiently fast in equilibrium, some smaller feasible innovations must be harmful. Moreover, the fact that the team is large ensures that innovations are produced sufficiently fast at the outset.%
	\footnote{\Cref{proposition:detrimental_n} is valid even if the stopping threshold cannot be reached with a single innovation: although  innovations have negligible ex-ante benefit at the outset (as they arrive arbitrarily fast), they can still be beneficial ex-post and, therefore, also harmful. See \Cref{equilibrium} for details.}


	I derive necessary and sufficient conditions for small increments in the stock to be harmful near the stopping threshold (\Cref{proposition:detrimental_longrun}). 
	This is guaranteed to occur after a sufficiently large reduction in the average frequency of innovations, provided their average size is increased just enough to keep the stopping threshold unchanged (\Cref{corollary:detrimental}).
	I also show that, under linear payoffs, innovations are always beneficial if they are sufficiently frequent; if innovations are instead sufficiently rare, and the team is sufficiently large, then only the last innovation produced in equilibrium can be beneficial (\Cref{proposition:linear_lambda}).
	Moreover, under linear payoffs, some innovations are harmful if the distribution of their size is sufficiently dispersed (\Cref{proposition:linear_M}).
	This is in spite of the fact that, under linear payoffs, increased dispersion induces uniformly higher effort and continuation payoff (\Cref{proposition:linear_cs}).
	To gain intuition for the effects of dispersion, suppose that the size of innovations is fixed and that effort stops after one innovation in the symmetric equilibrium. Then, making the size of innovations random (while preserving its mean) raises their expected benefit, since agents exert effort after a small enough draw. As a consequence, agents exert higher effort initially, are better off, and a mean-sized innovation is less beneficial (as it makes effort stop).

	The possibility of harmful innovations rests on two assumptions.
	First, the size of innovations must be random in order for them to be harmful: if the size of innovations is fixed, continuation payoffs increase over time in the symmetric equilibrium (\Cref{corollary:fixed_size}), since innovations are always beneficial ex-ante.
	Second, although increasing the stock has diminishing returns, in order for innovations to be harmful, the marginal cost of effort must increase as the stock grows: if the cost depends on effort alone, the continuation payoff in the symmetric equilibrium is increasing in the stock (\Cref{proposition:cost}).
	The logic behind this result is as follows: If the cost depends on effort alone, the marginal benefit of effort must fall as the stock grows in equilibrium, since effort is reduced. 
	This means that the marginal value of increasing the stock falls, on average, after each innovation produced in equilibrium. 
	Therefore, increasing the stock is always beneficial, since it is beneficial to do so after effort stops.



	Because innovations may be harmful in the symmetric equilibrium, agents who produce them may wish to delay their disclosure and adoption.
	To explore this possibility, I allow each agent to \emph{discard} the innovations that she produces, after observing their size.
	This yields a unique strongly symmetric equilibrium in which effort and continuation payoff are uniformly higher than in the symmetric equilibrium with forced disclosure (\Cref{theorem:disposal:equilibrium}).
	Moreover, if arbitrarily large innovations are feasible and the team is sufficiently large, then ex-ante payoffs in this equilibrium exceed those of all equilibria with forced disclosure (\Cref{proposition:disposal_welfare}).
	Thus, even though achieving the welfare benchmark requires all innovations to be disclosed, disposal occurs in equilibrium and is beneficial.
		%
		%

	The rest of the paper is organised as follows.
	I summarise the related literature in \Cref{literature}, describe the model in \Cref{model}, and derive the welfare benchmark (\Cref{proposition:benchmark}) in \Cref{benchmark}.
	In \Cref{equilibrium}, I characterise the symmetric Markov perfect equilibrium (\Cref{theorem:equilibrium}) and obtain \Cref{proposition:detrimental_n,proposition:detrimental_longrun,corollary:detrimental,corollary:fixed_size,proposition:cost}.
	In \Cref{linear}, I focus on linear payoffs to obtain \Cref{proposition:linear_lambda,proposition:linear_M,proposition:linear_cs}.
	\Cref{proposition:sse,proposition:ppe} are in \Cref{ppe}.
	In \Cref{disposal}, I analyse the game with disposal and obtain \Cref{theorem:disposal:equilibrium,proposition:disposal_welfare}. The conclusion is in \Cref{conclusion}.

	\subsection{Literature review}
	\label{literature}
	This paper contributes to the literature on dynamic games of public-good provision without a fixed goal (e.g.\ \textcite{fershtman1991} and \textcite{battaglini2014}).
	In these games, as in mine, agents exert costly effort to increase the stock of a public good, their flow payoffs rise as the stock grows, and returns fall. 
	My game has two distinguishing features: effort produces lumpy and randomly-sized increments,
	and the marginal cost of effort increases as the stock grows.%
	\footnote{
		\label{footnote:literature}
		To the best of my knowledge, only the model of \textcite{ferrari2017} incorporates either of these two features, and it encompasses both. 
		However, they only study Nash equilibria in open-loop strategies, and do not comment on the possibility of harmful increments.
		There are several studies of variants of the model of \textcite{fershtman1991} in which the stock evolves stochastically but \emph{continuously.} See e.g.\ \textcite{wang2010,kwon2022}.
	}
	Due to these features, increments produced in the symmetric Markov perfect equilibrium may decrease continuation payoffs, and allowing agents to discard them is beneficial. 

	This paper is also related to the strategic-experimentation literature, in particular to \textcite{bonatti2011}.
	In their model, agents exert hidden effort to obtain a breakthrough, and failed attempts make them more pessimistic about its feasibility.
	In the welfare benchmark as well as in the symmetric equilibrium, effort and continuation payoff fall over time before the breakthrough.
	In my model, effort becomes costlier and its returns fall as the stock grows, but flow payoffs rise. As a consequence, continuation payoffs rise over time in the benchmark, and may both rise or fall over time in equilibrium.
	Moreover, I allow agents to discard increments in the stock, a policy with no analogue in \textcite{bonatti2011}.

	Within the vast theoretical literature on innovation, this paper is most related to \textcite{cetemen2023}.
	They analyse a collective-search model in which discoveries accumulate over time, search has stochastic returns, and its cost rises as progress is made.
	Unlike in my model, each agent decides when to quit irreversibly, in order to exploit the best discovery to date.
	Moreover, discoveries are never harmful in their model, since they are arbitrarily small and frequent. Hence, agents have no incentive to discard them.

	Finally, this paper speaks to the literature studying which equilibrium payoffs can be sustained in dynamic partnership games with imperfect monitoring.
	\textcite{marx2000} and \textcite{horner2022} show that strongly symmetric equilibria (SSE) can sustain a higher payoff than the symmetric Markov perfect equilibrium (MPE) in dynamic games with \emph{perfect} monitoring of the aggregate contribution.
	\textcite{abreu1991} show that, in \emph{repeated} games with imperfect monitoring, efficient SSE exist if perfectly-revealing `bad-news' signals are available. 
	In my model --- a dynamic game with imperfect monitoring --- the symmetric MPE is the unique SSE, and it is inefficient.
	Moreover, this result is not driven by the lack of bad-news signals, since it would continue to hold (with the same proof) if past effort were revealed after every innovation.
	\section{Model and welfare benchmark}
	\label{model}
	In this section, I introduce the model and discuss the main assumptions.

	Time is continuous, indexed by $t \in \mathbb R_+$, and discounted at rate $r > 0$.
	There are $n \ge 2$ identical agents, indexed by $i$, and a public good.
	Its time-$t$ stock is $x_t \ge 0$.
	At time $t$, agent $i$ exerts effort $a_t^i \in [0,1]$ and receives a payoff flow $r [b(x_t)-c(a_t^i,x_t)]$, 
	for some $b: \mathbb{R}_+ \rightarrow \mathbb{R}_+$ and $c: [0,1] \times \mathbb{R}_+ \rightarrow \mathbb{R}_+$ such that $c(0,x) = 0$ for all $x$.
	Agents' efforts are hidden. 

	The stock $x_t$ takes some initial value $x_0$ and evolves as follows:
	Agent $i$ produces increments in $x_t$ at rate $\lambda a_t^i$, where $\lambda > 0$.
	Increments have random size $z \ge 0$, drawn from a continuous CDF $F$ with finite mean $\mu$ and convex support which includes $0$.%
	\footnote{
		\label{footnote:figure}
		To ease exposition, figures are drawn using an $F$ with binary support. The figures illustrate the effort schedules ($\alpha_e$ and $\alpha_d$) and the continuation-payoff functions ($v_e$ and $v_d$) in the equilibria of \Cref{theorem:equilibrium,theorem:disposal:equilibrium}. The proofs of the existence of these equilibria and of the uniqueness of $\alpha_e$, $\alpha_d$, $v_e$ and $v_d$ are valid for arbitrary distributions $F$ with finite means.
	}
		%
		%
	The arrival of increments, their size, and the identity of the agents inducing them, are public.
	Suppose without loss that $r = 1$.
	\footnote{This is equivalent to multiplying $\lambda$ by $1/r$.
	}
		%
		%

	Assume that $b$ and $c$ are twice continuously differentiable with locally-bounded derivatives, that $b$ is unbounded, and that $c(a,x)$ is increasing and convex in $a$.
	Suppose also that $b(x)-c(a,x)$ is increasing and concave in $x$.%
	\footnote{Terms such as `increasing' and `concave' are meant in the weak sense.}
	That is, keeping effort fixed, the payoff flow increases as the stock grows, but at a decreasing rate.
	Assume further that $c_{12}(a,x) > 0$ and that $c_{11}(a,x)$ is increasing in $x$, where subscripts denote partial derivatives.
	That is, the cost of effort becomes strictly steeper and weakly `more convex' as the stock grows.
	Suppose that $b'(0) > 0$ and $c_1(0,0) = 0$, so that exerting effort is optimal at the outset.
	Finally, assume that, as $x \rightarrow \infty$, either $b'(x) \rightarrow 0$ or $c_1(0,x) \rightarrow \infty$. This ensures that it is socially efficient to exert no effort if the stock is large enough.


	For some results, I will restrict attention to \emph{linear multiplicative} payoffs:
	\begin{equation}
		\label{eq:linear}
		b(x) = x \quad \& \quad c(a,x) = ax.
	\end{equation}

	I interpret the stock $x_t$ as the quality of a shared \emph{technology}, and increments in $x_t$ as \emph{innovations}.
	Given this interpretation, \eqref{eq:linear} may be understood as follows.
	In each period, agents can either use or (try to) improve the technology.
	Improving the technology ($a^i_t = 1$) yields no immediate payoff, and using it ($a^i_t = 0$) yields a payoff flow equal to its current quality, $x_t$.
	(Interpret $0 < a^i_t < 1$ as improving the technology with probability $a^i_t$ and using it with probability $1-a^i_t$.)

	\smallskip

	\noindent \textbf{Histories, strategies and solution concepts.}
	Since effort is hidden, a (public) \emph{history} is a sequence  
	\begin{equation}
		\label{eq:history_informal}
		h = (x_0,(t_1,z_1,i_1),\dots,(t_m,z_m,i_m))
	\end{equation}
	such that agent $i_1$ obtains an innovation of size $z_1$ at time $t_1$, agent $i_2$ one of size $z_2$ at time $t_2 > t_1$, and so on (with $m < \infty$, and $h = x_0$ being the initial history).
	The stock after the $m{\text{th}}$ innovation is
	\begin{equation*}
		x_h = x_0 + \sum_{\ell = 1}^m z_\ell.
	\end{equation*}
	Agents reach a new history whenever an innovation is produced.
	I restrict attention to pure public strategies.%
	\footnote{
		\label{footnote:pure}
		As I argue in \Cref{proof:pure}, mixed public strategies are essentially without loss of generality (by standard results), and pure public strategies involve no further loss. 
	}
	A (pure, public) \emph{strategy} $\sigma^i$ specifies an effort schedule $\sigma^i_h : (t_h,\infty) \rightarrow [0,1]$ for each history $h$, where $t_h = t_m$ if $h$ satisfies \eqref{eq:history_informal} and $t_h = 0$ if $h = x_0$.%
	\footnote{
		\label{footnote:measure}
		I require that the map $H_m \times (0,\infty) \to [0,1]$ given by $(h,s) \mapsto \sigma^i_h(t_h+s)$ be (Lebesgue) measurable for all $m \in \{0,1,\dots\}$, where $H_m$ is the set of histories of the form \eqref{eq:history_informal} for each $m > 0$ and $H_0 = \mathbb{R}_+$.} 
	Agent $i$ exerts effort $\sigma^i_h(t)$ at any time $t > t_h$ such that no innovation was produced within $[t_h,t)$.
	A strategy $\sigma^i$ is \emph{Markov} if effort is pinned down by the current stock;
	formally, if $\sigma^i_h(t) = \alpha^i(x_h)$ for all $h$ and $t > t_h$, given some $\alpha^i : \mathbb R_+ \rightarrow [0,1]$.
	In this case, I identify $\sigma^i$ with $\alpha^i$.

	If agents play a profile of (arbitrary) strategies $\sigma = (\sigma^i)_{i=1}^n$, agent $i$'s \emph{continuation payoff at} $h$ is 
	\begin{equation}
		\label{eq:payoff}
		v^i_\sigma(h) = \mathbb E\left(\sum_{\ell = 0}^{\tilde m} \int_{\tilde t_\ell}^{\tilde t_{\ell+1}} e^{t_h-t} \left[b\left(x_{\tilde h^\ell}\right) - c\left(\sigma^i_{\tilde h^\ell}(t),x_{\tilde h^\ell}\right)\right]\text dt \right)
	\end{equation}
	where $m \in \{0,1,\dots,\infty\}$ is the total number of innovations produced after time $t_h$, $h^\ell$ is the history reached after the $\ell\text{th}$ innovation (at time $t_\ell$) for each $0 < \ell \le m$, $h^0 = h$, $t_0 = t_h$, and $t_{m + 1} = \infty$ if $m < \infty$.

	A \emph{public perfect equilibrium} (PPE) is a profile of strategies $\sigma = (\sigma^i)_{i = 1}^n$ such that $\sigma^i$ is a best response for agent $i$ against $\sigma^{-i} = (\sigma^j)_{j \ne i}$ at any history $h$; that is, such that $\sigma^i$ maximises $v^i_{(\hat \sigma^i,\sigma^{-i})}(h)$ among all strategies $\hat \sigma^i$.
	A \emph{strongly symmetric equilibrium} (SSE) is a PPE $(\sigma^i)_{i = 1}^n$ such that $\sigma^i = \sigma^j$ for all $i$ and $j$.
	I identify any SSE with the strategy inducing it. 
	A symmetric \emph{Markov perfect equilibrium} (MPE) is a SSE that is Markov.

	By a slight abuse of notation, given a profile of Markov strategies $\alpha = (\alpha^i)_{i = 1}^n$ and $x \ge 0$, I write $v^i_\alpha(x)$ for agent $i$'s continuation payoff under $\alpha$ at any history $h$ such that $x_h = x$.
	It will be useful to express payoffs recursively as%
	\footnote{
			To obtain \eqref{eq:payoff_markov}, note that $z \mapsto v^i_\alpha(x+z)$ is $F$-integrable for all $x$, since $b(x)-c(a,x)$ is bounded below and concave in $x$.
			Then \eqref{eq:payoff} implies that 
			\begin{align*}
				v^i_\alpha(x) &= \int_0^t \left\{\left(1-e^{-s}\right)\left[b(x)-c\left(\alpha^i(x),x\right)\right] + e^{-s} \mathbb{E}_F[v^i_\alpha(x+\tilde z)] \right\}\lambda\sum_{j = 1}^n \alpha_j(x) e^{-\lambda s \sum_{j = 1}^n \alpha_j(x)}\text{d}s 
				\\& \quad + e^{-\lambda t\sum_{j = 1}^n \alpha_j(x)}\left\{\left(1-e^{-t}\right)\left[b(x)-c\left(\alpha^i(x),x\right)\right] + e^{-t}v^i_\alpha(x)\right\}
			\end{align*}
			for all $t > 0$. Differentiating with respect to $t$ and letting $t \to 0$ yields \eqref{eq:payoff_markov}.
	} 
	\begin{equation}
		\label{eq:payoff_markov}
		v^i_\alpha(x) = b(x) - c\left(\alpha^i(x),x\right) +  \lambda \left(\sum_{j = 1}^n\alpha^j(x)\right) \left\{\mathbb{E}_F\left[v^i_\alpha(x+\tilde z)\right]-v^i_\alpha(x)\right\}.
	\end{equation}
	The difference $b(x)-c(\alpha^i(x),x)$ is the current payoff flow, and the last term is the net expected future benefit. This benefit is given by the rate $\lambda \sum_{j = 1}^n \alpha^j(x)$ at which innovations are produced, multiplied by their expected value for agent $i$: the difference between her continuation payoff $v^i_\alpha(x+z)$ following an innovation of size $z$, and her current payoff $v^i_\alpha(x)$, weighted by the distribution $F$ of $z$.

	\smallskip

	\noindent \textbf{Discussion of the assumptions.}
	Effort determines the frequency of innovations but does not influence their size. 
	Allowing the latter would complicate the analysis but is likely to yield similar insights, provided effort not only increases the size of increments on average, but also its variance.

	Innovations are lumpy. 
	This is crucial for the analysis, and is justified by the agents' inability to observe their opponents' progress in real time. 

	The quality of the technology $x_t$ does not deteriorate over time. I use this assumption to obtain a unique SSE (\Cref{proposition:sse}) and to show that effort cannot be sustained beyond the stopping threshold in the single-agent problem (\Cref{proposition:ppe}). 
	Depreciation is not likely to qualitatively affect the analysis of the symmetric MPE: innovations would continue to be harmful in some cases, and allowing agents to discard them would be beneficial.

	In the literature on dynamic public-good games, the cost $c(a,x)$ of effort typically does not vary with the stock $x$.
	I rule out this case by imposing that $c_{12}(a,x) > 0$.
	If the cost of effort did not vary with the stock, all innovations would be beneficial in the symmetric MPE (\Cref{proposition:cost}).
	\subsection{Social-welfare benchmark}
	\label{benchmark}
	In this section, I describe how agents should behave in order to maximise welfare.
	The conclusions are in line with the literature, and all innovations are beneficial.

	Let $v_*(x)$ denote the highest average of agents' ex-ante payoffs achievable in the game with initial stock $x$.
	Note that it is efficient for agents to exert symmetric effort, since the cost of effort $c(a,x)$ and the rate of arrival of innovations $\lambda a$ are convex and linear in $a$, respectively.
	Then, in light of \eqref{eq:payoff_markov},
	\begin{equation}
		\label{eq:benchmark}
		v_*(x) = \max_{a \in [0,1]} \big\{b(x)- c(a,x) + a \lambda n \{\mathbb E_F[v_*(x+\tilde z)]-v_*(x)\}\big\}
	\end{equation}
	and a Markov strategy $\alpha$ is efficient (in the sense of inducing payoffs $v_*(x)$ at any stock $x$ if played by all agents) if and only if $a = \alpha(x)$ attains the maximum in \eqref{eq:benchmark} for each $x$.%
	\footnote{
		\label{footnote:benchmark}
		This follows from e.g.\ Theorem 3.1.2 of \textcite{piunovskiy2020}. This theorem guarantees that $z \mapsto v_*(x+z)$ is finite and $F$-integrable for all $x \ge 0$, so that the right-hand side of \eqref{eq:benchmark} is well-defined.
	}
	\begin{proposition}
		\label{proposition:benchmark}
	  Efficient effort $\alpha_*(x)$ is decreasing in the stock $x$ and ceases once $x$ reaches a threshold $x_* > 0$, which uniquely solves
	  \begin{equation}
			\label{eq:benchmark_cutoff}	
			\lambda n \{\mathbb{E}_F[b(x_*+\tilde z)]-b(x_*)\} = c_1(0,x_*).
		\end{equation}
	  The maximal payoff $v_*(x)$ is 
		increasing in $x$.
		\footnote{
			There may be multiple efficient Markov strategies. $\alpha_*$ is the pointwise smallest.
		}
	\end{proposition}
	The proof of \Cref{proposition:benchmark} is in \Cref{proof:benchmark}. 
	Decreasing effort 
	is standard in dynamic public-good games without a fixed goal. It is due to payoffs being concave, and to the cost of effort becoming steeper and more convex as $x$ grows.
	Condition \eqref{eq:benchmark_cutoff} states that the marginal social benefit of exerting effort at $x_*$ equals its marginal cost.
	The maximal payoff $v_*(x)$ is increasing in $x$ because flow payoffs are, keeping effort fixed.
	Hence, every innovation is \emph{beneficial} in the sense that it raises the continuation payoff.
	Under linear multiplicative payoffs \eqref{eq:linear}, efficient effort is `bang bang' and $x_* = \lambda\mu n$ (see \Cref{figure:benchmark}).
	%
	\begin{figure}[h]
		\begin{center}
			\begin{tikzpicture}
				\begin{groupplot}[
	      	group style={group size = 2 by 1},
					height = 6cm,
					width = 6cm,
					axis lines = left,
					ylabel style={rotate=-90},
					xtick={0,2,4},
					xticklabels={{},2,4},
					extra x ticks = {\xstar},
					extra x tick style={grid=major,tick label style={anchor=north}},
					extra x tick labels={$x_*$}, 
					extra y ticks = {0},
					extra y tick style={tick label style={anchor=north east}},
			    ]
			    \nextgroupplot[ymax = 1.1, ytick = {1}, ylabel = $\alpha_*(x)$]
						\addplot [domain = 0:\xstar,samples=100, thick] {1};
						\addplot [domain = \xstar:\xstarmax,samples=100, thick]{0};
			    \nextgroupplot[xshift = 1.5cm, ymin = 0, ylabel = $v_*(x)$,ytick={2,4}]
						\addplot [domain = 0:\xstar,samples=100, thick] 
						{
							((\l*\n)/(1 + \l*\n*\r)^2)*((1 - \r)*\e*e^((1 + \l*\n*\r)*(x - \xstar)/(\e*(1 + \l*\n))) + \r*(\xstar + (1 + \l*\n*\r)*x + \z))
						};
						\addplot [domain = \xstar:\xstarmax,samples=100, thick]{x};
					%
		    \end{groupplot}
			\end{tikzpicture}
			\caption{Effort (left) and continuation payoff (right) in the social-welfare benchmark. 
			In this figure, $n = 5$, $b(x) = x$, $c(a,x) = ax$, $\lambda = 10$, and $F$ assigns probability 0.99 to $z =0.01$ and probability 0.01 to $z = 5$. Effort ceases when the stock reaches $x_* = 2.995$.}
			\label{figure:benchmark}
		\end{center}
	\end{figure}
	\section{Symmetric equilibrium}
	\label{equilibrium}
	In this section, I characterise the unique symmetric MPE of the game (\Cref{theorem:equilibrium}).
	(We will see in \Cref{ppe} that this is also the unique SSE.)
	I show that some innovations are harmful in this equilibrium if the team is sufficient large (\Cref{proposition:detrimental_n}).
	I then derive necessary and sufficient conditions for small innovations to be harmful near the threshold at which effort stops (\Cref{proposition:detrimental_longrun}).
	This occurs after a large enough reduction of the average frequency of innovations, provided that their average size is increased just enough to keep the threshold unchanged (\Cref{corollary:detrimental}).
	I also argue that the hypotheses of innovations having random size and of the marginal cost of effort being increasing in the stock, are necessary for innovations to be harmful (\Cref{corollary:fixed_size} and \Cref{proposition:cost}, respectively).

	Given a Markov strategy $\alpha$ and $x \ge 0$, write $\hat v_\alpha(x)$ for the largest ex-ante payoff that an agent can achieve (across arbitrary strategies) in the game with initial stock $x$, if all her opponents play $\alpha$. 
	In light of \eqref{eq:payoff_markov}, 
	\begin{equation}
	  \label{eq:equilibrium}
		\hat v_\alpha(x) = \max_{a \in [0,1]} \big\{b(x)-c(a,x) + \lambda [a+(n-1)\alpha(x)]\{\mathbb E_F[\hat v_\alpha(x+\tilde z)]-\hat v_\alpha(x)\} \big\}
	\end{equation}
	and $\alpha$ is a symmetric MPE if and only if $a = \alpha(x)$ attains the maximum in \eqref{eq:equilibrium} for all $x$.%
	\footnote{
		\label{footnote:equilibrium}
		This follows from e.g.\ Theorem 3.1.2 of \textcite{piunovskiy2020}. This theorem guarantees that $z \mapsto \hat v_\alpha(x+z)$ is finite and $F$-integrable for all $x \ge 0$, so that the right-hand side of \eqref{eq:equilibrium} is well-defined. It also ensures that, given any Markov strategy $\alpha$, there exists a Markov strategy that is a best response, at any history and among arbitrary strategies, against all opponents playing $\alpha$.
	}
	The following result, proved in \Cref{proof:equilibrium}, characterises the unique symmetric MPE of the game.
	%
	\begin{theorem}
		\label{theorem:equilibrium}
	  There exists a unique symmetric MPE $\alpha_e$.
	  Effort $\alpha_e(x)$ is decreasing in the stock $x$, does not exceed the benchmark $\alpha_*(x)$, and ceases once $x$ reaches a threshold $x_e \in (0,x_*)$, which uniquely solves 
	  \begin{equation}
			\label{eq:equilibrium_cutoff}
			\lambda\{\mathbb{E}_F[b(x_e+\tilde z)] - b(x_e)\} = c_1(0,x_e).
		\end{equation}
	  Effort $\alpha_e(x)$ and continuation payoff $v_e(x)$ are continuously differentiable, except possibly at $x_e$ and $y_e = \inf\{x \ge 0 :\alpha_e(x) < 1\}$, and they are continuous. 
		%
	\end{theorem}
	Condition \eqref{eq:equilibrium_cutoff} is analogous to \eqref{eq:benchmark_cutoff}, with the left-hand side being the \emph{individual} benefit of effort instead of the social one.
	In light of \Cref{proposition:benchmark} (which is valid for $n = 1$), $x_e$ equals the stopping threshold of the single-agent problem. 
	The fact that effort in the symmetric MPE cannot be sustained above the single-agent threshold is standard in contribution games without depreciation. See e.g.\ \textcite{keller2005} and (Proposition 2 of) \textcite{battaglini2014}.
	We will see in \Cref{ppe} that, in this model, \emph{no} equilibrium can sustain effort beyond this threshold.

	\begin{figure}[h]
		\begin{center}
			\begin{tikzpicture}
				\begin{groupplot}[
	      	group style={group size = 2 by 1},
					height = 6cm,
					width = 6cm,
					axis lines = left,
					ylabel style={rotate=-90},
					xtick={0},
					xticklabels={},
					extra x ticks = {\yf,\xf},
					extra x tick style={grid=major,tick label style={anchor=north}},
					extra x tick labels={$y_e$,$x_e$},
					ytick={1},
					extra y ticks = {0},
					extra y tick style={tick label style={anchor=north east}},
			    ]
			    \nextgroupplot[ymax = 1.1, ytick = {1}, ylabel = $\alpha_e(x)$]
						\addplot[domain = 0:\yf,samples=100,thick]{1};
						\addplot[domain = \yf:\xf,samples=100,thick]
						{
							(-\e*(-1 + \r)*(-1 +e^((\r*(\e*\l*(-1 + \r) + x - \l*\r*\z))/\e) + \l*\r) + \r*(-x + \l*\r*\z))/(\l*(-1 + \n)*(\r^2)*x)
						};
						\addplot[domain = \xf:\xfmax,samples=100,thick]{0};
			    \nextgroupplot[xshift = 1.5cm, ylabel = $v_e(x)$, ymax = 1.4, ymin = 0]
						\addplot [domain = 0:\yf,samples=100,thick] 
						{	
							1.77667 - 0.505215*e^(2.94118*x) + 0.333333*x
						};
						\addplot [domain = \yf:\xf,samples=100,thick]
						{
							(1/(\l*\r))*((1/\r - 1)*\e*(e^(\r*(x - \xf)/\e) - 1) - (1 - \l*\r)*x + \xf)
						};
						\addplot[domain = \xf:\xfmax,samples=100,thick]{x};
					%
		    \end{groupplot}
			\end{tikzpicture}
			\caption{Effort (left) and continuation payoff (right) in the symmetric MPE. In this figure, $n = 5$, $b(x) = x$, $c(a,x) = ax$, $\lambda = 10$, and $F$ assigns probability 0.99 to $z =0.01$ and probability 0.01 to $z = 5$. The equilibrium and efficient stopping thresholds are $x_e = 0.599$ and $x_* = 2.995$, respectively. Effort becomes interior at $y_e \approx 0.19$.}
			\label{figure:equilibrium}
		\end{center}
	\end{figure}

	Effort $\alpha_e$ and continuation payoff $v_e$ are depicted in \Cref{figure:equilibrium} assuming that the distribution $F$ is binary and that flow payoffs are linear multiplicative \eqref{eq:linear}.
	Large innovations lead to effort stopping and are always beneficial.
	Small innovations, however, are \emph{harmful} (in the sense that they cause the continuation payoff to fall) unless the stock is close enough to $x_e$. 
	To understand why innovations can be harmful, note first that they have two opposing effects on the continuation payoff: 
	they increase flow payoffs, but induce agents to exert lower effort, and this slows down the growth of flow payoffs in the future.
	In spite of this trade-off, in the welfare benchmark, innovations are always beneficial, since agents exert the socially-optimal amount of effort.
	In equilibrium, however, each agent neglects the social value of her own effort, and innovations are harmful in some cases.

	The next result shows that, in large enough teams, harmful innovations are produced with positive probability in the equilibrium with initial stock $x = 0$.
	Its proof is in \Cref{proof:detrimental}.
	\begin{proposition}
		\label{proposition:detrimental_n}
		If $n$ is sufficiently large, then $v_e$ is not increasing on $[0,x_e]$, where $x_e$ is given by \eqref{eq:equilibrium_cutoff}.
	\end{proposition}
	The logic behind \Cref{proposition:detrimental_n} is as follows.
	For any stock level $x$, the expected continuation payoff after an innovation (i.e. $\mathbb{E}_F[v_e(x+\tilde z)]$) remains bounded as $n$ grows large, since the stopping threshold $x_e$ does not vary with $n$.
	Moreover, the hypothesis that $c_1(0,0) = 0$ implies that aggregate effort at the outset, viz.\ $n\alpha_e(0)$, grows without bound as $n$ does. 
	Then, the ex-ante benefit of an innovation at $x = 0$, namely
	\begin{equation*}
		\mathbb{E}_F[v_e(\tilde z)]-v_e(0) = \frac{\mathbb{E}_F[v_e(\tilde z)]-b(0)+c(\alpha_e(0),0)}{1+\lambda n \alpha_e(0)},
	\end{equation*}
	vanishes in the limit as $n$ grows large.
	Now, the assumption that $c_{12}(0,x) > 0$ implies that individual effort $\alpha_e(x)$ vanishes, for any given $0 < x < x_e$, as $n$ diverges, so that 
	\begin{equation}
		\label{eq:d_approx}
		c_{12}(0,x) \approx \frac{\mathrm{d}}{\mathrm{d}x} c_1(\alpha_e(x),x) = \frac{\mathrm{d}}{\mathrm{d}x}\lambda \{\mathbb{E}_F[v_e(x+\tilde z)]-v(x)\}
	\end{equation}
	when $n$ is large, where the equality follows from the first-order condition for $\alpha_e(x)$ derived from \eqref{eq:equilibrium}.%
	\footnote{I show as part of the proof of \Cref{theorem:equilibrium} that $\mathbb{E}_F[v_e(x+\tilde z)]$ is continuously differentiable in $x$ (\Cref{lemma:dv} in \Cref{proof:equilibrium}).}
	The assumption that $c_{12}(0,0) > 0$ then rules out the possibility that, for large $n$, $v_e$ is approximately constant on the support of $F$ (which includes $0$).
	Therefore, innovations at $x = 0$ must be harmful with positive probability if $n$ is sufficiently large.

	The next result characterises when a small increment immediately below the stopping threshold $x_e$ (given by \eqref{eq:equilibrium_cutoff}) is harmful in the symmetric MPE. Its proof is in \Cref{proof:detrimental}.
	\begin{proposition}
		\label{proposition:detrimental_longrun}
		The left derivative of payoffs $v_e(x)$ at the stopping threshold $x_e$ exists and is given by   
		\begin{equation}
			\label{eq:d_longrun}
			v'_e(x^-_e) = b'(x_e) - (n-1)c_1(0,x_e)\frac{\lambda \{b'(x_e)-\mathbb E_F[b'(x_e + \tilde z)]\} + c_{12}(0,x_e)}{c_{11}(0,x_e) + \lambda (n-1)c_1(0,x_e)}.
		\end{equation}
	\end{proposition}
	A small increment in the stock below $x_e$ increases each agent's flow payoff (first term in \eqref{eq:d_longrun}), but also reduces her net expected future benefits (second term), since it induces opponents to exert lower effort. $(n-1)c_1(0,x_e)$ approximates the net expected future benefit per unit of effort exerted by each opponent (namely $\lambda (n-1) \{\mathbb{E}_F[v_e(x+\tilde z)] - v_e(x)\}$) near $x_e$.%
	\footnote{This follows from \eqref{eq:equilibrium_cutoff} since $v_e(x)$ is continuous and equals $b(x)$ on $[x_e,\infty)$.} 
	The fraction equals the absolute value of the left derivative of effort $\alpha_e(x)$ at $x_e$.
	Effort falls both because the increment decreases its marginal benefit (first term in the numerator), since $b$ is concave, 
	and because the increment raises the marginal cost of effort (second term in the nominator). 
	The fall in effort is mitigated by the curvature of the cost (captured by $c_{11}(0,x_e)$) and by the fact that effort choices at a given stock level are strategic substitutes across agents. 
	This last effect is captured by the term $\lambda (n-1)c_1(0,x_e)$ in the denominator.

	By \Cref{proposition:detrimental_longrun}, $v'_e(x^-_e)$ is decreasing in $n$ (and strictly so unless $c_{11}(0,x_e) = 0$). That is, a small increment in the stock immediately below $x_e$ becomes less beneficial (or more harmful) as the team becomes larger, in line with \Cref{proposition:detrimental_n}. This is because the increment causes a larger drop in the aggregate effort exerted by opponents, but it increases flow payoffs by the same amount (as the cutoff $x_e$ does not vary with $n$).

	Under linear multiplicative payoffs \eqref{eq:linear}, $v_e'(x^-_e) = 1-1/\lambda$, so that small increments immediately below $x_e$ are beneficial if and only if innovations are sufficiently frequent. This is the case in the example of \Cref{figure:equilibrium}, since $\lambda = 10$.
	Intuitively, such increments cause a smaller drop in effort if innovations are more frequent, since the aforementioned substitution effect is larger (as effort is more productive), while the magnitude of all other effects is unchanged.
	The next result has a similar flavour, but holds for arbitrary payoffs. Its proof is in \Cref{proof:detrimental}.
	\begin{corollary}
		\label{corollary:detrimental}
		Let $\mathcal{F} = (F_{\lambda'})_{\lambda' > 0}$ be an inversely FOSD-ordered family of distributions satisfying the model assumptions, and $\hat x > 0$ be such that 
		\begin{equation}
			\label{eq:equilibrium_cutoff_lambda}
			\lambda'\left\{\mathbb{E}_{F_{\lambda'}}[b(\hat x+\tilde z)] - b(\hat x)\right\} = c_1(0,\hat x)
		\end{equation}
		for all $\lambda' > 0$.
		There exists $\lambda^* \in \mathbb{R}_+$ such that,%
		\footnote{The value of $\lambda_*$ depends on the choice of $b$, $c$, $n$, $\mathcal{F}$ and $\hat x$, but not on the value of $\lambda$.} 
		assuming that $F = F_{\lambda}$, $v'_e(x^-_e) < \mathrel{(>)} 0$ if $\lambda < \mathrel{(>)} \lambda^*$.
		Moreover, $\lambda^* > 0$ provided
		\begin{equation}
			\label{eq:detrimental_lambda_n}
			n > \frac{c_{11}(0,\hat x)b'(\hat x)}{c_1(0,\hat x)c_{12}(0,\hat x )}+1.
		\end{equation}
	\end{corollary}
	To understand \Cref{corollary:detrimental}, note first that for any $\lambda' < \mathrel{(>)} \lambda$, there exists a distribution $F'$ satisfying the model assumptions that first-order stochastically dominates (resp.\ is dominated by) $F$, and such that replacing $\lambda$ by $\lambda'$ and $F$ by $F'$ preserves the marginal benefit of effort at the stopping threshold $x_e$:  
	\begin{equation}
		\label{eq:cutoff_marg_ben}
		\lambda \{\mathbb{E}_F[b(x_e+\tilde z)] - b(x_e)\} = \lambda' \{\mathbb{E}_{F'}[b(x_e+\tilde z)] - b(x_e)\}.
	\end{equation}
	\Cref{corollary:detrimental} delivers a threshold $\lambda^*$ such that $v'_e(x^-_e) < \mathrel{(>)} 0$ after this transformation if $\lambda' < \mathrel{(>)} \lambda^*$.%
	\footnote{
		\label{footnote:corollary}
		In particular, the combination of \eqref{eq:equilibrium_cutoff_lambda} and of the assumption that $F = F_\lambda$ is equivalent to \eqref{eq:cutoff_marg_ben} holding with $F' = F_{\lambda'}$ for all $\lambda' > 0$. This equivalence follows from \eqref{eq:equilibrium_cutoff}.
	}
	This means that small increments immediately below the threshold $x_e$ are beneficial if innovations are small and frequent, but they are harmful if innovations are large and rare.


	The possibility of harmful innovations in the symmetric MPE relies on two assumptions.
	First, innovations must have random size in order to be harmful.
	More specifically:
	\begin{corollary}
		\label{corollary:fixed_size}
		Suppose that the distribution $F$ is degenerate instead of satisfying the assumptions in \Cref{model}.
		If all other assumptions in \Cref{model} are maintained, (the proof of) \Cref{theorem:equilibrium} remains valid and continuation payoffs are increasing over time in the symmetric MPE.
		%
	\end{corollary}
	\Cref{corollary:fixed_size} follows readily from the fact that, in the symmetric MPE, all innovations are beneficial ex-ante; that is, $\mathbb{E}_F[v_e(x+\tilde z)] \ge v_e(x)$ for all $x$. This inequality is clear if no effort is exerted at $x$, since $b$ is increasing, and follows from \eqref{eq:equilibrium} otherwise.

	Second, although payoffs $b(x)-c(a,x)$ are concave in the stock $x$, the marginal cost $c_1(a,x)$ must increase as $x$ grows for innovations to be harmful.
	\begin{proposition}
		\label{proposition:cost}
		Suppose that the cost of effort $c(a,x)$ does not vary with $x$ instead of satisfying $c_{12}(a,x) > 0$ and $c_1(0,0) = 0$.
		Suppose also that all other assumptions in \Cref{model} are maintained, that $b$ is strictly concave and
		\begin{equation}
			\label{eq:outset}
			\lambda \{\mathbb{E}_F[b(\tilde z)]-b(0)\} > c_1(0,0) > 0.
		\end{equation} 
		Then \Cref{theorem:equilibrium} remains valid and $v_e(x)$ is increasing in $x$.%
		\footnote{The additional hypotheses in \Cref{proposition:cost} are imposed to ensure that \eqref{eq:equilibrium_cutoff} admits a unique solution $x_e >0$.
	}
	\end{proposition}
	The proof of \Cref{proposition:cost} is in \Cref{proof:detrimental}. 
	To understand why the result holds, suppose that the cost does not vary with the stock and note that, in equilibrium, the marginal cost of effort $c_1(\alpha_e(x),x)$ falls as the stock $x$ grows, since effort $\alpha_e(x)$ falls (by \Cref{theorem:equilibrium}).
	Then, the marginal benefit of effort $\lambda \{\mathbb{E}_F[v_e(x+\tilde z)]-v_e(x)\}$ decreases on the interval where effort $\alpha_e(x)$ is interior. As a consequence, for any $x$ in this interval, the marginal value of increasing the stock $x$ falls, on average, after an innovation; that is, $v_e'(x) \ge \mathbb{E}_F[v_e'(x+\tilde z)]$.
	Iterating the inequality implies that this value exceeds the expected marginal benefit of increasing the long-run stock; that is, $v_e'(x) \ge \mathbb{E}_{G^x}[b'(\tilde y)]$ where \setword{$G^x$}{word:Gx} is the distribution of $x + \sum_{\ell = 1}^{\tilde m} \tilde z_\ell$, $m = \min\{\ell \ge 1: x + z_1 + \dots + z_\ell \ge x_e\}$, and $(z_\ell)_{\ell = 1}^\infty$ are independent draws from $F$.
	Because $b$ is increasing, increasing the stock $x$ is beneficial: $v_e'(x) \ge 0$.%
	\footnote{This argument establishes that $v_e$ is increasing on $[y_e,x_e]$. Monotonicity on $[0,y_e]$ then follows from a similar argument; see \Cref{proof:detrimental} for details.}
	\subsection{Linear multiplicative payoffs}
	\label{linear}

	In this section, I obtain additional results about the symmetric MPE assuming linear multiplicative payoffs \eqref{eq:linear}. 
	I show that innovations are always beneficial if they are sufficiently frequent, and that only the last innovation that is produced in equilibrium can be beneficial if innovations are sufficiently rare and the team is sufficiently large (\Cref{proposition:linear_lambda}). I also show that some innovations are harmful if the distribution of their size is sufficiently dispersed (\Cref{proposition:linear_M}).
	This is in spite of the fact that increased dispersion raises both effort and continuation payoff at all stock levels (\Cref{proposition:linear_cs}).

	Note that effort ceases at $x_e = \lambda\mu$ in the symmetric MPE if \eqref{eq:linear} holds, by \eqref{eq:equilibrium_cutoff}.
	Then $M_F(\lambda\mu-x)$ innovations are produced on average if the initial stock is $x < \lambda\mu$, where $M_F(\Delta) = \mathbb E\left[\min \left\{m \ge 1 : \sum_{\ell = 1}^m \tilde z_\ell \ge \Delta \right\}\right]$
	is the expected number of increments needed to increase the stock by $\Delta > 0$, and $(z_\ell)_{\ell = 1}^\infty$ are independent draws from $F$.
	Recall from \Cref{theorem:equilibrium} that $y_e$ is the stock level at which effort becomes interior.
	\begin{lemma}
		\label{lemma:linear}
		If \eqref{eq:linear} holds, then $v'_e(x) = 1-M_F(\lambda\mu-x)/\lambda$ for all $y_e < x < \lambda\mu$.
	\end{lemma}
	The proof of \Cref{lemma:linear} is in \Cref{proof:linear}. \Cref{lemma:linear} implies that $v_e'$ is increasing on $(y_e,\lambda\mu)$ (since $M_F(\Delta)$ is increasing in $\Delta$, clearly); that is, a marginal increment in the stock is more valuable the closer the stock is to the stopping threshold $\lambda\mu$. Intuitively, this is because the increment raises flow payoffs by the same amount, by \eqref{eq:linear}, but causes fewer innovations to be delayed in the future.%
	\footnote{This intuition is not valid if the initial stock lies below $y_e$, since in this case innovations are delayed only after the current stock exceeds this threshold.}

	\begin{proposition}
		\label{proposition:linear_lambda}
		Suppose that \eqref{eq:linear} holds. 
		Then, given any $0 < \hat x < \lambda\mu$, $v_e$ is strictly decreasing on $[\hat x,\lambda\mu]$ if $\lambda \le 1$ and $n$ is sufficiently large.
		Moreover, $v_e$ is (globally) increasing if $\lambda$ is sufficiently large.
	\end{proposition}
	 
	The proof of \Cref{proposition:linear_lambda} is in \Cref{proof:linear}. The first part of this result implies that, given any $0 < \hat x < \lambda\mu$, all but the last innovation produced in the symmetric MPE with initial stock $\hat x$ are harmful if $\lambda \le 1$ and $n$ is large enough. The continuation payoff $v_e$ is decreasing on $[\hat x,\lambda\mu]$ in this case since $v_e'$ is increasing on $(y_e,\lambda\mu)$ (by \Cref{lemma:linear}), $v_e'(x)$ is negative in the limit as $x$ tends to $\lambda\mu$ from below (by \Cref{proposition:linear_lambda}, since $\lambda \le 1$) and, as I show in the proof of \Cref{proposition:linear_lambda}, $y_e < \hat x$ for large enough $n$.

	The second part of \Cref{proposition:linear_lambda} states that all innovations are beneficial if they arrive sufficiently frequently. 
	Intuitively, this is because the randomness of the size of innovations plays a limited role in this case (due to the law of large numbers), and innovations are beneficial if their size is fixed (\Cref{corollary:fixed_size}).
	\begin{proposition}
		\label{proposition:linear_M}
		Suppose that \eqref{eq:linear} holds and that 
		\begin{equation}
			\label{eq:linear_onlyif}
			\frac 1 {\mu\left(1-\frac 1n\right)} \int_0^{\lambda\mu\left(1-\frac 1n\right)} F(z) \,\mathrm{d}z > \lambda-1
		\end{equation}
		Then $v_e$ is not increasing on $[0,\lambda\mu]$. 
	\end{proposition}
	The proof of \Cref{proposition:linear_M} is in \Cref{proof:linear}.
	This result states that some innovations are harmful in the symmetric MPE if the CDF $F$ is sufficiently `dispersed', in the sense that it assigns enough mass to the lower end of its support, keeping its mean $\mu$ fixed.%
	\footnote{Note also that \eqref{eq:linear_onlyif} holds if $\lambda \le 1$, in line with \Cref{proposition:linear_lambda}, and if $n$ is sufficiently large (provided $F(\lambda\mu) <1$), in line with \Cref{proposition:detrimental_n}.}
	Intuitively, given any initial stock $x < \lambda\mu$, the average number of innovations produced in equilibrium is large if $F$ is dispersed in the aforementioned sense, so that a small innovation at $x$ is harmful (by \Cref{lemma:linear}), unless $y_e > x$. 
	Moreover, an innovation raising the stock from $y_e$ to $\lambda\mu$ is harmful if $y_e$ is close to $\lambda\mu$, since effort is maximal at $y_e$ and stops at $\lambda\mu$, while efficient effort is maximal on $[0,\lambda\mu n)$.
	We shall see that, in spite of this, increasing the dispersion of $F$ (in the appropriate sense) is beneficial in the symmetric MPE.
	\begin{definition}
		\label{def:dispersion}
		A CDF $F^\dag$ is \emph{more dispersed} than $F$ if $F^\dag$ and $F$ have the same mean $\mu$ and $F^\dag(z) \ge F(z)$ for all $z \le \lambda\mu$.
	\end{definition} 
	If $F^\dag$ is more dispersed than $F$, then $F^\dag$ may be obtained from $F$ by decreasing the size of some draws, and increasing others that are large enough to make effort stop (i.e.\ larger than $\lambda\mu$), while keeping the average size unchanged.

	\begin{figure}[h]
		\begin{center}
			\begin{tikzpicture}
				\begin{groupplot}[
	      	group style={group size = 2 by 1},
					height = 6cm,
					width = 6cm,
					axis lines = left,
					ylabel style={rotate=-90},
					xtick={},
					xticklabels={},
					extra x ticks = {\xf},
					extra x tick style={grid=major,tick label style={anchor=north}},
					extra x tick labels={$x_e$},
					extra y ticks = {0},
					extra y tick style={tick label style={anchor=north east}},
			    ]
			    \nextgroupplot[ymax = 1.1, ytick = {1}, ylabel = $\alpha_e(x)$]
						\addplot[domain = 0:\yf,samples=100,thick]{1};
						\addplot[domain = \yf:\xf,samples=100,thick]
						{
							(-\e*(-1 + \r)*(-1 +e^((\r*(\e*\l*(-1 + \r) + x - \l*\r*\z))/\e) + \l*\r) + \r*(-x + \l*\r*\z))/(\l*(-1 + \n)*(\r^2)*x)
						};
						\addplot[domain = \yfmed:\xf,samples=100,thick,dashed]
						{
							(0.231713517215 + 0.0289802883568*e^(3.34899328859*x) - 0.746492985972*x)/x
						};
						\addplot[domain = \yflow:\xf,samples=100,thick,dotted]
						{
							(0.17080186425 + 0.000902982262593*e^(6.74324324324*x) - 0.370741482966*x)/x
						};
						\addplot[domain = \xf:\xfmax,samples=100,thick]{0};
			    \nextgroupplot[xshift = 1.5cm, ylabel = $v_e(x)$, ytick={1}, ymax = 1.4, ymin = 0]
						\addplot [domain = 0:\yf,samples=100,thick] 
						{	
							1.77667 - 0.505215*e^(2.94118*x) + 0.333333*x
						};
						\addplot [domain = \yfmed:0,samples=100,thick, dashed] 
						{	
							1.05227647908 - 0.149256292983*e^(5.24411106725*x) + 0.626097867001*x
						};
						\addplot [domain = \yflow:0,samples=100,thick, dotted] 
						{	
							0.6635 - 0.0431584926944*e^(8.57180710122*x) + 0.771251931994*x
						};
						\addplot [domain = \yf:\xf,samples=100,thick]
						{
							(1/(\l*\r))*((1/\r - 1)*\e*(e^(\r*(x - \xf)/\e) - 1) - (1 - \l*\r)*x + \xf)
						};
						\addplot [domain = \xf:\yfmed,samples=100, thick, dashed, dash phase=2pt]
						{
							(1/(\l*\rmed))*((1/\rmed - 1)*\e*(e^(\rmed*(x - \xf)/\e) - 1) - (1 - \l*\rmed)*x + \xf)
						};
						\addplot [domain = \xf:\yflow,samples=100, thick, dotted, dash phase=2.2pt]
						{
							(1/(\l*\rlow))*((1/\rlow - 1)*\e*(e^(\rlow*(x - \xf)/\e) - 1) - (1 - \l*\rlow)*x + \xf)
						};
						\addplot[domain = \xf:\xfmax,samples=100,thick]{x};
						\addplot[domain = 0:\xf,samples=100,opacity=0.25]{x};
		    \end{groupplot}
			\end{tikzpicture}
			\caption{Effort (left) and continuation payoff (right) in the symmetric MPE, for varying distributions $F$ ordered by dispersion. The solid curves correspond to the parameter values of the previous figures; in particular $F$ assigns probability $0.99$ to $z = 0.01$ and $0.01$ to $z = 5$ (and is the most dispersed). For the dashed curve, $F$ assigns probability $p \approx 0.96$ to $z = 0.01$ and $1-p$ to $z =1.5$; for the dotted curve, $F$ assigns probability $p' \approx 0.93$ to $z = 0.01$ and $1-p'$ to $z = 0.75$ (and is the least dispersed). The values of the remaining parameters are as in the previous figures. The threshold $x_e = 2.995$ does not vary with $F$ since the mean $\mu$ is fixed. The gray line through the origin in the right panel corresponds to the myopic payoff $b(x) = x$.}
			\label{figure:equilibrium_dispersion}
		\end{center}
	\end{figure}
	\begin{proposition}
		\label{proposition:linear_cs}
		Let $F^\dag$ satisfy the assumptions in \Cref{model},%
		\footnote{That is, the CDF $F^\dag$ is continuous with convex support in $\mathbb{R}_+$ containing $0$.} and be more dispersed than $F$.
		If \eqref{eq:linear} holds, then effort $\alpha^\dag_e$ and continuation payoff $v^\dag_e$ in the symmetric MPE under $F^\dag$ are no lower than their analogues under $F$. That is, $\alpha^\dag_e(x) \ge \alpha_e(x)$ and $v^\dag_e(x) \ge v_e(x)$ for all $x \ge 0$.
	\end{proposition}
	\Cref{proposition:linear_cs} is proved in \Cref{proof:linear}, and the logic behind it is as follows:
	On one hand, decreasing the size $z$ of an innovation cannot lower payoffs $v_e(x+z)$ more than one-to-one (and may in fact increase them), since, as \Cref{figure:equilibrium_dispersion} suggests, the gap between equilibrium payoffs $v_e(x)$ and myopic payoffs $b(x)=x$ shrinks as the stock grows.%
	\footnote{The fact that $v_e-b$ is decreasing is a typical feature of dynamic public-good games without a fixed goal, and does not rely on \eqref{eq:linear} in this model. See \Cref{proposition:equilibrium:general} in \Cref{proof:equilibrium}.}
	On the other hand, increasing the size of an innovation that causes effort to stop raises equilibrium payoffs one-to-one. 
	Because $F^\dag$ may be obtained from $F$ via a sequence of such operations, drawing the first innovation from $F^\dag$ instead of $F$ increases the expected payoff in the continuation equilibrium; that is, $\mathbb{E}_{F^\dag}[v_e(x+\tilde z)] \ge \mathbb{E}_F[v_e(x+\tilde z)]$.
	As a consequence, agents exert more effort before the first innovation if it is drawn from $F^\dag$ instead of $F$, and their ex-ante payoff is higher.
	Repeating this reasoning yields that effort and ex-ante payoff rise further if \emph{all} innovations are drawn from $F^\dag$ instead of from $F$.

	%
	%
	\subsection{Other equilibria}
	\label{ppe}
	In this section, I show that the symmetric MPE is the unique SSE of the game (\Cref{proposition:sse}).
	I also show that, in any PPE, the stock reaches its steady-state when it attains the single-agent stopping threshold (\Cref{proposition:ppe}).

	Recall from \Cref{model} the definition of SSE.
	\begin{proposition}
		\label{proposition:sse}
		The symmetric MPE $\alpha_e$ is the unique SSE.
	\end{proposition}
	The multiplicity results of \textcite{marx2000,lockwood2002,horner2022} suggest that \Cref{proposition:sse} hinges on the assumption that monitoring is imperfect.
	If aggregate effort were observable, or could be perfectly inferred from the trajectory of the stock, multiple SSE would exist and some would induce higher payoffs than the symmetric MPE.
	The proof of \Cref{proposition:sse} (in \Cref{proof:ppe}) follows a `backward-induction' logic and relies on the fact that effort must vanish as the stock grows arbitrarily large.

	Recall from \Cref{model} the definition of PPE.
	The game admits many asymmetric PPE, some of which yield payoff gains over the symmetric MPE, due to the agents' ability to coordinate.
	However, in any PPE, no effort is exerted once the stock exceeds the single-agent stopping threshold.
	This follows from the next result, which also establishes that effort in any PPE stops completely only after the single-agent threshold is reached. 
	\begin{proposition}
		\label{proposition:ppe}
		In any PPE, some effort is exerted after time $t$ if $x_t < x_e$ and only if $x_t \le x_e$, where $x_e$ is given by \eqref{eq:equilibrium_cutoff}.%
		\footnote{
			\textcite{keller2005} prove an analogous result for Markov equilibria with `finite switching'.
			\Cref{proposition:ppe} is also similar to Theorem 1 in \textcite{gueron2015}, who considers a discrete-time setting.
		}
	\end{proposition}
	The proof of \Cref{proposition:ppe} is in \Cref{proof:ppe}.
	To see why no effort is exerted above $x_e$ in any PPE $\sigma$, note first that no effort is exerted at any history $h$ with stock $x_h > x_*$, or else average continuation payoffs would exceed $b(x_h)$, contradicting \Cref{proposition:benchmark}.
	Consider now the supremum $\hat x$ of $x_h$ across histories $h$ at which some effort is exerted under $\sigma$.
	At histories such that $x_h$ is arbitrarily close to $\hat x$, it is arbitrarily likely that the first innovation brings the stock above $\hat x$ and causes effort to stop.
	Then, exerting effort at $h$ is (at best) approximately as profitable as exerting the same amount of effort at $x_h$ in the single-agent problem, since effort is hidden. 
	This implies that $\hat x \le x_e$, since effort is strictly suboptimal at any stock $x > x_e$ in the single-agent problem. 

	\Cref{proposition:ppe} implies that the highest welfare across all PPE with initial stock $x$ need not approximate the benchmark $v_*(x)$ as agents become arbitrarily patient (i.e.\ as $\lambda$ diverges). That is, the folk theorem need not hold.
	To see why, suppose that payoffs are linear and multiplicative \eqref{eq:linear}, so that $\alpha_*(x) = \mathbf{1}_{x < \lambda \mu n}$ and $x_e = \lambda\mu$.
	In this case, subject to no effort being exerted above $\lambda\mu$, average ex-ante payoffs are maximised (regardless of the initial stock) if all agents adopt the Markov strategy $\bar \alpha(x) = \mathbf{1}_{x \le \lambda\mu}$.
	Moreover, the payoff $\bar v(x)$ induced by this strategy profile satisfies $\lim\sup_{\lambda \rightarrow \infty} \bar v(x)/v_*(x) < 1$ if the size of innovations is exponentially-distributed (i.e., if $F(z) = 1-e^{-z / \mu}$).%
	\footnote{
		More specifically, $\bar v(x)/v_*(x) \rightarrow e^{1-1/n}/n < 1$ as $\lambda \rightarrow \infty$ in this case,  since $v_*(x) = \lambda \mu n e^{\frac{x/\mu-\lambda n}{1+\lambda n}}$ and $\bar v(x) = \frac{\lambda n \mu(1+\lambda)}{1+\lambda n} e^{\frac{x/\mu-\lambda}{1+\lambda n}}$ for all $x \le \lambda \mu$. The expressions for $v_*(x)$ and $\bar v(x)$ are easily obtained using the results in Section 2.2 of \textcite{polyanin2008}.
	}
	\section{Disposal}
	\label{disposal}
	In this section, I extend the model by allowing each agent to freely dispose of the innovations that she produces, after observing their size.
	This raises effort and continuation payoff at all stock levels in the SSE, which remains unique and Markov (\Cref{theorem:disposal:equilibrium}).
	Under natural conditions, all agents are ex-ante better off in this equilibrium than in any PPE with forced disclosure (\Cref{proposition:disposal_welfare}).

	\subsection{Model}

	\label{disposal:model}

	I enrich the model as follows:
	Whenever an agent obtains an innovation, she immediately decides whether to disclose it or discard it, after observing its size.
	Disclosure raises the stock by the size of the innovation and reveals the identity of the innovator, as in the baseline model. 
	Disposal is not observed by other agents.
	I constrain agents to play pure public strategies.
		%
		%

	We recover the baseline model (described in \Cref{model}) by restricting strategies so that all innovations are disclosed.
	I refer to the baseline model as the game \emph{with forced disclosure,} and to this model as the game \emph{with disposal}.

	\smallskip

	\noindent \textbf{Histories, strategies and solution concepts.} 
	Each agent reaches a new private history whenever she produces or discards an innovation, or any agent discloses one.
	Public histories are defined as in the baseline model (see \eqref{eq:history_informal}), but they now only record disclosed innovations.
	A (pure, public) strategy $\xi^i = (\sigma^i,\chi^i)$ specifies, for each public history $h$, an effort schedule $\sigma^i_h : (t_h,\infty) \rightarrow [0,1]$ and a disclosure policy $\chi^i_h : (t_h,\infty) \times \mathbb{R}_+ \rightarrow \{0,1\}$.%
	\footnote{
		The measurability restriction on $\sigma^i$ is unchanged (see footnote \ref{footnote:measure}). I require that $(h,s,z) \mapsto \chi^i_h(t_h + s,z)$ be a (Lebesgue) measurable map $H_m \times (0,\infty) \times \mathbb{R}_+ \to \{0,1\}$ for all $m \in \{0,1,\dots\}$.
	}
	As before, agent $i$ exerts effort $\sigma^i_h(t)$ at any time $t > t_h$ such that no innovation was disclosed within $[t_h,t)$.
	Moreover, if agent $i$ produces an innovation of size $z$ at such a time $t$, she discloses it if $\chi^i_h(t,z) = 1$ and discards it otherwise.
	A strategy $\xi^i = (\sigma^i,\chi^i)$ is \emph{Markov} if there exists a pair $\pi^i = (\alpha^i,\delta^i)$ of maps $\alpha^i : \mathbb R_+ \rightarrow [0,1]$ and $\delta^i : \mathbb R_+^2 \rightarrow \{0,1\}$ such that $\sigma^i_h(t) = \alpha^i(x_h)$ and $\chi^i_h(t,z) = \delta^i(x_h,z)$ for all histories $h$, times $t > t_h$, and $z \ge 0$.
	In this case, I identify $\xi^i$ with $\pi^i$.

	If agents play a profile of (arbitrary) strategies $\xi = (\sigma^i,\chi^i)_{i = 1}^n$, agent $i$'s continuation payoff $v^i_\xi(h)$ at $h$ may be expressed as the right-hand side of \eqref{eq:payoff}, with the histories $h^\ell$ obviously having a different joint distribution.

	A \emph{strongly symmetric equilibrium} (SSE) is a profile of strategies $\xi = (\xi^i)_{i = 1}^n$ such that $\xi^i = \xi^j$ for all $i$ and $j$, and $\xi^i$ is a best response for each agent $i$ against $\xi^{-i} = (\xi^j)_{j \ne i}$ at any public history $h$. The latter requirement means that $\xi^i$ maximises $v^i_{(\hat \xi^i,\xi^{-i})}(h)$ among all strategies $\hat \xi^i$.
	I identify any SSE with the strategy that induces it.
	A symmetric \emph{Markov perfect equilibrium} (MPE) is a SSE that is Markov.

	Given any profile of Markov strategies $\pi = (\alpha^i,\delta^i)_{i = 1}^n$ and $x \ge 0$, I write $v^i_\pi(x)$ for agent $i$'s continuation payoff at any public history $h$ with stock $x_h = x$.
	The analogue of \eqref{eq:payoff_markov} is 
	\begin{equation}
		\label{eq:payoff_disposal_markov}
		v^i_\pi(x) = b(x) - c(a,x) + \lambda \sum_{j = 1}^n \alpha^j(x) \mathbb{E}_F\left[\delta^j(x,\tilde z) \left(v^i_\pi(x+\tilde z)-v^i_\pi(x)\right)\right].
	\end{equation}
	The only difference relative to \eqref{eq:payoff_markov} is that the last term features the disclosure decision $\delta^j(x,z)$.
	\subsection{Equilibrium}
	\label{disposal:analysis}
	Given a Markov strategy $\pi = (\alpha,\delta)$ and $x \ge 0$, write $\hat v_\pi(x)$ for the largest ex-ante payoff that an agent can achieve (across arbitrary strategies) in the game with initial stock $x$, if all her opponents play $\pi$. 
	Let $\mathcal{D}$ be the set of all Lebesgue measurable $D \subseteq \mathbb{R}_+$.
	In light of \eqref{eq:payoff_disposal_markov}, 
	\begin{align}
		\nonumber
		\hat v_\pi(x) &= b(x) + \max_{\substack{a \in [0,1] \\ D \in \mathcal{D}}} \big\{a \lambda \mathbb E_F\left[\mathbf{1}_{\tilde z \in D}\left(\hat v_\pi(x+\tilde z)-\hat v_\pi(x)\right)\right] -c(a,x) \big\}
		\\\label{eq:disposal:equilibrium}
		&\quad + \lambda (n-1)\alpha(x) \mathbb E_F\left[\delta(x,\tilde z)\left(\hat v_\pi(x+\tilde z)-\hat v_\pi(x)\right)\right]
	\end{align}
	and $\pi$ is a symmetric MPE if and only if $a = \sigma(x)$ and $D = \{z \in \mathbb{R}_+ : \delta(x,z) = 1\}$ achieve the maximum in \eqref{eq:disposal:equilibrium} for all $x$.%
	\footnote{This follows from e.g.\ Theorem 46.18 of \textcite{davis2018}. This theorem guarantees that $z \mapsto \hat v_\pi(x+z)$ is finite and $F$-integrable for all $x$, and that, given any $\pi$, there exists a Markov strategy that is a best response, at any public history and among arbitrary strategies, against all opponents playing $\pi$.
	}

	Recall the symmetric MPE $\alpha_e$ of the game with forced disclosure (\Cref{theorem:equilibrium}).
	The following result characterises a symmetric MPE of the game with disposal, and shows that it is essentially the only SSE.
	The proof is in \Cref{proof:disposal}.
	\begin{theorem}
		\label{theorem:disposal:equilibrium}
		The game with disposal admits a symmetric MPE $(\alpha_d,\delta_d)$.
		Effort $\alpha_d$ and continuation payoff (labelled $v_d$) are no lower than their analogues in the game with forced disclosure: $\alpha_d(x) \ge \alpha_e(x)$ and $v_d(x) \ge v_e(x)$ for all $x$.
		If, for some $\hat x$, $v_e$ is increasing on $[\hat x,\infty)$, then all innovations are disclosed and the equilibria coincide on this interval: $\delta_d(x,z) = 1$ and $\alpha_d(x) = \alpha_e(x)$ for all $x \ge\hat x$ and $z \ge 0$.
		If $v_e$ is not increasing on $[\hat x,\infty)$, then $v_d(\hat x) > v_e(\hat x)$ and $v_d$ is not increasing on $[\hat x,\infty)$.
		Moreover, any SSE $(\sigma,\chi)$ of the game with disposal satisfies $\sigma = \alpha_d$ and induces continuation payoff $v_d(x_h)$ after any public history $h$.
	\end{theorem}
	Although there may exist multiple SSE, \Cref{theorem:disposal:equilibrium} shows that effort and continuation payoff are uniquely pinned down.%
	\footnote{See \Cref{proposition:sse_general} in \Cref{proof:ppe} for a complete characterisation of SSE. Multiplicity arises since agents may be indifferent between adopting and discarding certain innovations, and any way of breaking ties yields an equilibrium.}
	Moreover, as I show in \Cref{proof:disposal}, $\alpha_d(x)$ inherits the qualitative properties of $\alpha_e(x)$: it is continuous and decreasing in $x$, weakly lower than the benchmark $\alpha_*(x)$, and vanishes once $x$ reaches the threshold $x_e$, given by \eqref{eq:equilibrium_cutoff}.
	\begin{figure}[h!]
		\begin{center}
			\begin{tikzpicture}
				\begin{groupplot}[
	      	group style={group size = 2 by 1},
					height = 6cm,
					width = 6cm,
					axis lines = left,
					ylabel style={rotate=-90},
					xtick={0,\xf},
					xticklabels={{}, $x_e$},
					extra x ticks = {\ydp,\xfhatp},
					extra x tick style={tick label style={anchor=north}},
					extra x tick labels={$y_\dag$, $x_\dag$},
					extra y ticks = {0},
					extra y tick style={tick label style={anchor=north east}},
			    ]
			    \nextgroupplot[ymax = 1.1, ytick = {1}, ylabel = $\alpha_d(x)$]
						\draw[white,fill=gray,opacity=0.25] (axis cs: \ydp, 0) rectangle (axis cs: \xfhatp, 1.1);
						
						\addplot[domain = 0:\ydp,samples=100,thick]{1};
						\addplot[domain = \ydp:\xfhatp,samples=100,thick]
						{
							(\z/x - 1/\lp)/(\n - 1)
						};
						\addplot[domain = \xfhatp:\xfp,samples=100,thick]
						{
							(\ep*(-1 + e^((x - \lp*(\ep + \z))/\ep) + \lp) - x + \lp*\z)/(\lp*(-1 + \n)*x)
						};
						\addplot[domain = \yfp:\xfp,samples=100,dotted, thick]
						{
							(\ep*(-1 + e^((x - \lp*(\ep + \z))/\ep) + \lp) - x + \lp*\z)/(\lp*(-1 + \n)*x)
						};
						\addplot[domain = \xfp:\xfmax,samples=100,thick]{0};
			    \nextgroupplot[ymin = 0, ymax = 2, xshift = 1.5cm, ylabel = $v_d(x)$, ytick={1,2}]

						\draw[white,fill=gray,opacity=0.25] (axis cs: \ydp, 0) rectangle (axis cs: \xfhatp, 2);
						\addplot [domain = 0:\yfp,samples=100,dotted, thick] 
						{
							1.77778 - 0.517277*e^(3*x) + 0.333333*x
						};
						\addplot [domain = \yfp:\xfp,samples=100,dotted, thick]
						{
							(\ep*(e^((x - \xfp)/\ep) - 1) - (1 - \lp)*x + \xfp)/\lp
						};
						\addplot[domain = 0:\ydp,samples=100,thick]
						{
							1.77778 - 0.0380577*e^(3*x) + 0.333333*x
						};
						\addplot[domain = \ydp:\xfhatp,samples=100,thick]
						{
							\z + (1 - 1/\lp)*x
						};
						\addplot[domain = \xfhatp:\xfp,samples=100,thick]
						{
							(\ep*(e^((x - \xfp)/\ep) - 1) - (1 - \lp)*x + \xfp)/\lp
						};
						\addplot[domain = \xfp:\xfmax,samples=100,thick]{x};
					%
		    \end{groupplot}
			\end{tikzpicture}
			\caption{
				Effort (left) and continuation payoff (right) in the symmetric MPE of the game with disposal. The dotted curves are effort $\alpha_e(x)$ (left) and continuation payoff $v_e(x)$ (right) in the symmetric MPE with forced disclosure.
				Parameter values are as in \Cref{figure:benchmark,figure:equilibrium}. 
				In particular, innovations have size $0.01$ or $5$.
				Large innovations are always disclosed, and small innovations are disclosed unless the (pre-disclosure) stock lies within the interval $(y_\dag,x_\dag)$, where $y_\dag \approx 0.36$ and $x_\dag \approx 0.49$.
				}
			\label{figure:disposal}
		\end{center}
	\end{figure}

	How does the ability to discard innovations affect the continuation payoff in the symmetric MPE $(\alpha_d,\delta_d)$? 
	First, it ensures that the continuation payoff increases over time.
	This is because agents have common preferences regarding which innovations to disclose and which to discard, and discard those that would (strictly) decrease payoffs. 
	That is, they discard an innovation that would raise the stock from $x$ to $x + z$ if and only if $v_d(x) > v_d(x+z)$.%
	\footnote{In particular, $\delta_d$ breaks ties in favour of disclosure. This feature delivers the full-disclosure result in \Cref{theorem:disposal:equilibrium}. 
	} 
	In the example of \Cref{figure:disposal}, innovations have two sizes ($\underline z = 0.01$ and $\bar z = 5$).
	Large innovations are always disclosed, and lead to effort stopping. Small innovations are discarded if and only if $x$ lies within an interval $(y_\dag,x_\dag)$, where $0 < y_\dag < x_\dag < x_e$. Hence, the stock evolves as follows: if its initial value $x$ does not exceed $y_\dag$, it increases in increments of size $\underline z$ until it exceeds $y_\dag$, then increases once more, by $\bar z$; if instead $y_\dag < x < x_\dag$ then the stock only increases once, to $x + \bar z$; finally, if $x_\dag \le x < x_e$, the stock increases in increments of size $\underline z$ or $\bar z$ until it reaches $x_e$ (and a single innovation of size $\bar z$ suffices for this to occur).

	Second, allowing the disposal of innovations does not ensure that the continuation payoff $v_d(x)$ is increasing in the stock $x$. 
	Indeed, innovations are discarded in the equilibrium $(\alpha_d,\delta_d)$ only if $v_d(x)$ is not increasing in $x$, since it is beneficial to adopt an innovation that raises the stock from $x$ to $x+z$ whenever $v_d(x+z) \ge v_d(x)$.
	Moreover, $v_d$ is increasing on an open interval containing $x_e$ if and only if $v_e$ is.%
	\footnote{This follows from \Cref{theorem:disposal:equilibrium}, since $\alpha_d(x) = 0$ for all $x \ge x_e$.}
	Intuitively, disposal does not alter the effect of a small increment in the stock near $x_e$, since innovations (drawn from $F$) are beneficial with high probability when the stock is close to $x_e$.

	Third, allowing the disposal of innovations increases the continuation payoff in the symmetric MPE, and strictly so unless future innovations are guaranteed to be beneficial.
	To see why, note that allowing agents to discard the \emph{first} innovation produced in the symmetric MPE increases the expected continuation payoff following the innovation (from $\mathbb{E}[v_e(x+\tilde z)]$ to $\mathbb{E}_F[\max\{v_e(x+\tilde z),v_e(x)\}]$, where $x$ is the initial stock). As a consequence, agents exert more effort before the first innovation if they may discard it, and have a higher ex-ante payoff.
	Repeating this argument implies that effort and ex-ante payoff rise further if agents may discard \emph{every} innovation.

	Under what conditions does the symmetric MPE with disposal induce higher ex-ante payoffs than any PPE with forced disclosure?
	To obtain a sufficient condition, label $G^x$ the distribution of $x + \sum_{\ell = 0}^{\tilde m} \tilde z_\ell$, where $z_0 = 0$, $(z_\ell)_{\ell = 1}^\infty$ are independent draws from $F$, $m = \min\{\ell \ge 0: x + z_0 + \dots + z_\ell > x_e\}$, and $x_e$ is given by \eqref{eq:equilibrium_cutoff}.%
	\footnote{Since $F$ is continuous and $\alpha_e(x) = 0$ for all $x \ge x_e$, this definition of $G^x$ coincides with the definition on page \pageref{word:Gx}, which assumed that $0 < \alpha_e(x) < 1$.}
	By \Cref{proposition:ppe}, individual ex-ante payoffs in any PPE of the game with forced disclosure are at most $\mathbb{E}_{G^{x}}[b(\tilde y)]$, where $x$ is the initial stock.
	The next result provides conditions for the ex-ante payoff in $(\alpha_d,\delta_d)$ to exceed this bound. Its proof is in \Cref{proof:disposal}. 
	\begin{proposition}
		\label{proposition:disposal_welfare}
		Suppose that
		\begin{equation}
			\label{eq:disposal_welfare}
			\lambda \mathbb{E}_F[\max\{b(x+\tilde z) - \mathbb{E}_{G^x}[b(\tilde y)],0\}] > c_1(0,x)
		\end{equation}
		and that $n$ is sufficiently large. Then $v_d(x) > \mathbb{E}_{G^x}[b(\tilde y)]$.
		In particular, at the initial stock is $x$, all agents are ex-ante better off in the symmetric MPE with disposal $(\alpha_d,\delta_d)$ than in any PPE of the game with forced disclosure.
	\end{proposition}
	\Cref{proposition:disposal_welfare} implies that, if \eqref{eq:disposal_welfare} holds, then the ability to discard innovations yields larger gains (in large teams) than the ability to coordinate.
	Moreover, if $v_d(x) > \mathbb{E}_{G^x}[b(\tilde y)]$ then disposal enables agents to achieve higher expected long-run payoffs, since these do no exceed $\mathbb{E}_{G^x}[b(\tilde y)]$ among PPE with forced disclosure, by \Cref{proposition:ppe}.

	Note that \eqref{eq:disposal_welfare} holds if $F$ has unbounded support and $x = 0$, since $c_1(0,0)=0$.
	It also holds under linear multiplicative payoffs \eqref{eq:linear} provided that $x < \lambda \mu$ and that $F$ satisfies the following property: keeping the mean $\mu$ fixed and given some $z_0 > 0$, $\mu_0 = \mathbb{E}_F[\tilde z | \tilde z \le z_0]$ is small and, fixing $\mu$ and $\mu_0$, $F(z_0)$ is close to $1$.%
	\footnote{
		\eqref{eq:disposal_welfare} holds under these assumptions since, given $\mu$, $\mu_0$ and $z_0$, $\mathbb{E}_{G^x}[b(\tilde y)]$ is bounded by some $E \in \mathbb{R}_+$ among all $F$ with the same $\mu$ and $\mu_0$ (as $\mu_0 > 0$), so that
	\begin{align*}
		\lambda\mathbb{E}_F[\max\{b(x + \tilde z) - \mathbb{E}_{G^x}[b(\tilde y)],0\}] &\ge \lambda\{\mathbb{E}_F[\mathbf{1}_{\tilde z > z_0}(x+\tilde z)] - [1-F(z_0)]E\} 
		\approx \lambda\mu > x = c_1(0,x),
	\end{align*} 
	where the approximate equality holds since $\mu_0$ is small and, given $\mu$ and $\mu_0$, $F(z_0)$ is close to $1$.
	} 
	These conditions imply that $F$ has a thick tail ($\mathbb{E}_F[\tilde z|\tilde z > z_0]$ is large) and that, with high probability, all innovations produced in the symmetric MPE with forced disclosure are `small' (of size less than $z_0$).

	Condition \eqref{eq:disposal_welfare} guarantees that the ability to discard innovations induces strong incentives to exert effort at $x$.
	More specifically, it ensures that the marginal benefit of effort at $x$ in the equilibrium $(\alpha_d,\delta_d)$, namely $\lambda \mathbb{E}_F[\max\{v_d(x+\tilde z)-v_d(x),0\}]$,
	is bounded away from its marginal cost $c_1(0,x)$, unless $v_d(x) > \mathbb{E}_{G^x}[b(\tilde y)]$.%
	\footnote{This is because, if $v_d(x) \le \mathbb{E}_{G^x}[b(\tilde y)]$ then $\lambda \mathbb{E}_F[\max\{v_d(x+\tilde z)-v_d(x),0\}] \ge \lambda \mathbb{E}_F[\max\{b(x+\tilde z) - \mathbb{E}_{G^x}[b(\tilde y)],0\}]$, since $v_d \ge b$.}

	\section{Conclusion}
	\label{conclusion}
	In this paper, I have studied the collective-action problem of improving a shared technology.
	Assuming that innovations are lumpy and randomly-sized, and that the opportunity cost of improving the technology rises as progress is made, small innovations may be socially-harmful in equilibrium. 
	As a consequence, allowing each agent to discard the innovations that she produces is beneficial.

	All results would remain true if agents may conceal innovations instead of discarding them, but may not secretly refine the improvements that they hide.
	Indeed, the equilibrium I constructed (in \Cref{theorem:disposal:equilibrium}) would survive since it is stationary (so that delaying disclosure is unprofitable), and no other equilibrium would arise since public strategies rule out disclosing previously-concealed innovations.
	Allowing agents to secretly refine concealed innovations would be one interesting avenue for future research.
	\begingroup
	\setlength\bibitemsep{0pt}
	\printbibliography
	\endgroup
	%
	\begin{appendices}
	\setcounter{section}{0}
	\crefalias{section}{app}
	\crefalias{subsection}{app}
	\section{Pure public strategies are without loss}
	\label{proof:pure}
	In this appendix, I argue that restricting attention to pure public strategies involves essentially no loss of generality. I first note that restricting attention to \emph{mixed} public strategies involves essentially no loss (\Cref{remark:pure}), then show that restricting attention to pure strategies involves no further loss (\Cref{lemma:pure}).

	Consider an approximation of the game introduced in \Cref{model} in which time is discrete and has length $\text{d}t \in (0,1/\lambda)$, time-$t$ effort $a^i_t$ is restricted to lie in $\{0,1/k,\dots,1\}$ for some $k \in \{1,2,\dots\}$, each agent $i$ produces an innovation at time $t$ with probability $\lambda a^i_t\text{d}t$ (and none otherwise), and $F$ has finite support.
	Note that this game has a `product structure' and no `short-run' players, in the sense of \textcite{fudenberg1994b}.
	Note also that their Theorem 5.2 (which concerns repeated games) easily extends to dynamic games with finite action spaces and finite-support state transitions, including to the game just described.
	Therefore:
	\begin{remark}
		\label{remark:pure}
		For any sequential equilibrium of the game just described, there exists a mixed-strategy public perfect equilibrium (PPE) inducing the same ex-ante payoff for each player.
	\end{remark}
	Consider now the game of \Cref{model} and recall the definition of a pure public strategy.
	A \emph{mixed} public strategy is a family $\rho^i = (\rho^{i,\beta})_{\beta \in [0,1]}$ of pure strategies; agent $i$ observes the realisation $\beta$ of a uniform draw on $[0,1]$ and plays the pure strategy $\rho^{i,\beta}$.%
	\footnote{
		Labelling $\Sigma$ the set of all pure public strategies, I require that the map $[0,1] \to \Sigma$ given by $\beta \mapsto \rho^{i,\beta}$ be measurable with respect to the $\sigma$-algebra on $\Sigma$ generated by sets of the form $\{\sigma \in \Sigma: \sigma_h(t) \in A\}$ where $h$ is a history, $t > t_h$, and $A \subset [0,1]$ is Borel.
		This definition is standard; see e.g.\ \textcite{bonatti2017} and references therein. In particular, $\pi^i_h$ and $\sigma^i$ are well defined in the proof of \Cref{lemma:pure} below.
	}

	\begin{lemma}
		\label{lemma:pure}
		For any mixed-strategy PPE, there exists a pure-strategy PPE inducing the same ex-ante payoff for each player.
	\end{lemma}

	\begin{proof}
		Fix a mixed-strategy PPE $\rho = (\rho^i)_{i = 1}^n$. Given any history $h$ and any agent $i$, let $\pi^i_h$ be the distribution over the draw $\beta \in [0,1]$ determining agent $i$'s behaviour, conditional on history $h$ having been reached.
		Define the pure strategy $\sigma^i$ by 
		\begin{equation}
			\label{eq:purification}
			\sigma^i_h(t) = \frac{\int \rho^{i,\beta}_h(t)  e^{-\lambda \int_{t_h}^t \rho^{i,\beta}_h} \text d \pi^i_h(\beta)}{\int e^{-\lambda \int_{t_h}^t \rho^{i,\beta}_h} \text d \pi^i_h(\beta)}
		\end{equation}
		for all $h$ and $t > t_h$, and note that $1-e^{-\lambda \int_{t_h}^t \sigma^i_h} =  1- \int e^{- \lambda \int_{t_h}^t \rho^{i,\beta}_h} \text d \pi^i_h(\beta)$. 
		That is, the earliest time after $t_h$ at which agent $i$ produces an innovation (assuming no other agent does) has the same distribution whether agent $i$ implements the effort schedule $\sigma^i_h$ at $h$, or observes the realisation $\beta$ of a draw from $\pi^i_h$ and implements $\rho^{i,\beta}_h$.
		Then, the highest payoff that any agent $i$ can achieve after $h$, across all public strategies, is preserved if her opponents play $(\sigma^i)_{j \ne i}$ instead of $(\rho^j)_{j \ne i}$.
		Hence, it suffices to show that $(\sigma^i)_{i = 1}^n$ is a PPE. By \Cref{proposition:ppe:dp} in \Cref{proof:ppe}, it suffices to show that \eqref{eq:ppe} holds for all $h$, $i$, and a.e.\ $t > t_h$. 
		
		Fix $i$ and $h$.
		Since $\rho$ is a PPE, $a = \rho^{i,\beta}_h(t)$ maximises the objective in \eqref{eq:ppe} for $\pi^i_h$-a.e.\ $\beta$ and a.e.\ $t > t_h$, by \Cref{proposition:ppe:dp}.
		Then, for a.e.\ $t > t_h$, there exist $\underline a,\bar a \in [0,1]$ such that $\underline a$ and $\bar a$ both maximise the objective in \eqref{eq:ppe} and $\underline a \le\sigma^i_h(t) \le \bar a$, so that \eqref{eq:ppe} holds, since $c(a,x_h)$ is convex in $a$.
	\end{proof}
	\section{Proof of \Cref{proposition:benchmark}}
	\label{proof:benchmark}
	I state and prove a lemma, then prove \Cref{proposition:benchmark}.

	\begin{lemma}
		\label{lemma:benchmark}
		$v_*-b$ is decreasing.
	\end{lemma}
	\begin{proof}[Proof of \Cref{lemma:benchmark}]
		Given a Markov strategy $\alpha$, let $v_\alpha(x)$ be the continuation payoff at stock $x$ if all agents play $\alpha$.
		Given $\epsilon > 0$, define the Markov strategy $\hat\alpha(x) = \alpha(x + \epsilon)$ and, given $x \ge \epsilon$, let $(x_t)_{t > 0}$ be the trajectory of the stock induced by $\alpha$ starting at the initial stock $x$.
		Note that  
		\begin{align*}
			&v_\alpha(x)-v_{\hat\alpha}(x-\epsilon) 
			= \mathbb{E}\left[\int_0^\infty e^{-t} \{b(\tilde x_t)-c(\alpha(\tilde x_t),\tilde x_t) \right.
			\\& \hspace{4.5cm} -\left[b(\tilde x_t-\epsilon) - c(\alpha(\tilde x_t),\tilde x_t - \epsilon)\right]\}\text{d}t\bigg]
			\le b(x)-b(x-\epsilon)
		\end{align*}
		where the inequality holds since $c_{12}(a,x) \ge 0$, $c(0,x) = 0$, and $b$ is concave.
		Since $v_{\hat\alpha}(x-\epsilon) \le v_*(x-\epsilon)$, considering a sequence of $\alpha$ such that $v_\alpha(x)$ converges to $v_*(x)$ yields $v_*(x) - v_*(x-\epsilon) \le b(x)-b(x-\epsilon)$, i.e.\ $v(x)-b(x) \le v(x-\epsilon)-b(x-\epsilon)$. 
		Hence, $v_*-b$ is decreasing.
	\end{proof}

	\begin{proof}[Proof of \Cref{proposition:benchmark}]
		It is clear that $v_*$ is increasing, since $b(x)-c(a,x)$ is increasing in $x$.
		Let $\alpha_*(x)$ be the least maximiser $a$ in \eqref{eq:benchmark} for each $x$.
		Note that $\alpha_* : \mathbb{R} \to [0,1]$ is measurable since $v_*$ is increasing.
		To show that $\alpha_*$ is decreasing, note that $\alpha_*(x)$ is a maximiser of the map 
		\begin{equation*}
			a \mapsto \frac{b(x)-c(a,x) + a\lambda n \mathbb{E}_F[v_*(x+\tilde z)]}{1 + a\lambda n}
		\end{equation*}
		as $v_*$ satisfies \eqref{eq:benchmark}, and that the derivative of this map has the same sign as 
		\begin{equation*}
			%
			%
			\gamma^*(a,x) = \mathbb{E}_F[v_*(x+\tilde z)] - [b(x)-c(a,x)] - \left(\frac1{n\lambda}+a\right)c_1(a,x).
		\end{equation*}
		Then, it suffices to show that $\gamma^*$ is decreasing in $a$ and in $x$.
		The former holds since $c(a,x)$ is increasing and convex in $a$.
		For the latter, note that $\mathbb{E}_F[v_*(x+\tilde z)] - b(x)$ is decreasing in $x$, by \Cref{lemma:benchmark}, since $b$ is concave; 
		note also 
		that $c(a,x)-ac_1(a,x) = -\int_0^a \int_{\hat a}^{a} c_{11}(\cdot ,x) \text{d}\hat a$ is decreasing in $x$ since $c_{11}(a,x)$ is increasing in $x$, and that $-c_1(a,x)/(n\lambda)$ is decreasing in $x$ since $c_{12}(a,x) \ge 0$.
		Hence, $\alpha_*$ is decreasing.

		To show that \eqref{eq:benchmark_cutoff} admits a unique solution $x_* > 0$, note that the solutions to \eqref{eq:benchmark_cutoff} are the roots of the map $\psi : \mathbb{R}_+ \to \mathbb{R}$ given by $\psi(x) = n\lambda\{\mathbb{E}_F[b(x+\tilde z)] - b(x)\} -c_1(0,x)$.
		Moreover, $\psi$ is continuous and strictly decreasing since $b$ is concave and $c_{12}(0,x) > 0$.
		Finally, $\psi(0) > 0$ and $\lim_{x \to \infty}\psi(x) < 0$ since $b'(0) > 0 = c_1(0,0)$ and, as $x \to \infty$, either $b'(x) \to 0$ or $c_1(0,x) \to \infty$. Then, $\psi$ has a unique root $x_*$ in $\mathbb{R}_+$, and $x_* >0$.

		It remains to show that $\alpha_*(x) = 0$ for $x \ge x_*$ and that $\alpha_*(x) > 0$ for $x < x_*$.
		The former holds since 
		\begin{equation*}
			\lambda n \{\mathbb{E}_F[v_*(x+\tilde z)] - v_*(x)\} \le \lambda n \{\mathbb{E}_F[b(x+\tilde z)]-b(x)\} \le c_1(0,x_*)
		\end{equation*}
		for $x \ge x_*$, where the first inequality follows from \Cref{lemma:benchmark} and the second holds since $b$ is concave.
		For the latter, suppose by means of contradiction that $\alpha_*(x) = 0$ for some $x < x_*$.
		Then, $v_* = b$ on $[x,\infty)$ since $\alpha_e$ is decreasing.
		Hence, $\alpha_*(x)$ cannot achieve the maximum in \eqref{eq:benchmark}, since $\psi(x) > \psi(x_*) = 0$.
	\end{proof}
	\section{Proof of \Cref{theorem:equilibrium}}
	\label{proof:equilibrium}
	I state two auxiliary results: \Cref{proposition:equilibrium:general,lemma:dv}.
	Together, these results immediately imply \Cref{theorem:equilibrium}.
	I prove \Cref{proposition:equilibrium:general} in \Cref{proof:equilibrium:general} and \Cref{lemma:dv} in \Cref{proof:dv}.

	\Cref{proposition:equilibrium:general} requires several definitions.
	Let $V$ be the set of (Lebesgue) measurable $v : \mathbb{R}_+ \to \mathbb{R}$ such that $b \le v \le v_*$.
	Note that there exists a unique $p : \mathbb{R}_+ \times (b(0),\infty) \to \mathbb{R}$ such that
	\begin{equation*}
		%
		%
		p(x,\ell) \in \arg\max_{a \in [0,1]} a \lambda \frac{\ell - [b(x)- c(p(x,\ell),x)]}{1+\lambda n p(x,\ell)} - c(a,x)
	\end{equation*}
	for all $x \ge 0$ and $\ell > 0$.
	Indeed, the objective is continuously differentiable and concave in $a$, and its derivative has the same sign as $\gamma(p(x,\ell),x,\ell)$, where
	\begin{equation}
		\label{eq:gamma}
		\gamma(a,x,\ell) = \ell - \left[b(x)-c(a,x)\right] -\left(\frac 1\lambda+na\right)c_1(a,x)
	\end{equation}
	is decreasing in $a$ and strictly so on $\{a \in [0,1] : \gamma(a,x,\ell) \le 0\}$.

	Let $\Gamma : [0,1] \times \mathbb{R}_+^2 \to \mathbb{R}$ be given by 
	\begin{equation*}
		\Gamma(a,x,\ell) = \frac{b(x)-c(a,x) + a \lambda n \ell}{1 + a \lambda n}.
	\end{equation*}
	Given $v : \mathbb{R}_+ \to \mathbb{R}$ such that $z \mapsto v(x+z)$ is $F$-integrable for all $x \ge 0$, define 
	\begin{equation}
		\label{eq:L}
		L_e v(x) = \mathbb{E}_F[v(x+\tilde z)] \quad \text{and} \quad L_d v(x) = \mathbb{E}_F[\max\{v(x),v(x+\tilde z)\}]
	\end{equation}
	for all $x \ge 0$.
	For any $v \in V$, let $A_kv : \mathbb{R}_+ \to [0,1]$ and $P_k v : \mathbb R_+ \rightarrow \mathbb R$ be given by
	\begin{equation*}
		A_kv(x) = p(x,L_kv(x)) \quad \text{and} \quad P_k v(x) = \Gamma\left(A_kv(x),x,L_kv(x)\right).
	\end{equation*}
	Recall from \Cref{disposal:model} that I refer to the baseline model as the game \emph{with forced disclosure,} and to the model introduced in \Cref{disposal:model} as the game \emph{with disposal}.
	\begin{proposition}
		\label{proposition:equilibrium:general}
		The game with forced disclosure admits a unique symmetric MPE $\alpha_e$, and $\alpha_e$ induces continuation payoff $v_e = \lim_{m \to \infty}(P_e)^mb$; the game with disposal admits a symmetric MPE $(\alpha_d,\delta_d)$ inducing continuation payoff $v_d = \lim_{m \to \infty}(P_d)^m b$, where $\delta_d(x,z) = 1$ if and only if $v_d(x+z) \ge v_d(x)$.%
		\footnote{In particular, $(P_d)^mb$ and $(P_e)^mb$ lie in $V$ for each $m \ge 0$, and both limits exist.}
		Moreover, for each $k \in \{d,e\}$, $\alpha_k = A_kv_k$, $P_kv_k = v_k$, $v_k-b$ is decreasing, $v_k$ is continuous, $\alpha_k$ is continuous and decreasing, and $\alpha_k \le \alpha_*$.
		Finally, $\alpha_k$ is strictly positive on $[0,x_e)$ and vanishes on $[x_e,\infty)$, where $x_e \in (0,x_*)$ uniquely solves \eqref{eq:equilibrium_cutoff}.
	\end{proposition}
	In order to state \Cref{lemma:dv}, let
	\begin{align*}
		%
		d(a,x) &= \frac{c_{11}(a,x)[b'(x)-c_2(a,x)] - (n-1)c_1(a,x)c_{12}(a,x)}
		{c_{11}(a,x)(1 + \lambda na) + \lambda(n-1)c_1(a,x)}
		\\ 
		g(a,x) &= \frac{\lambda[nac_{11}(a,x) + (n-1)c_1(a,x)]}
		{c_{11}(a,x)(1 + \lambda na) + \lambda(n-1)c_1(a,x)}
	\end{align*}
	for any $0 \le a \le 1$ and $x > 0$, noting that $c_1(a,x) > 0$. Set $d_e(x) = d(\alpha_e(x),x)$ and $g_e(x) = g(\alpha_e(x),x)$.
	Recall that $y_e = \inf\{x \ge 0 : \alpha_e(x) < 1\}$, and note that $y_e < x_e$, by \Cref{proposition:equilibrium:general}. Adopt the convention $\prod_{\ell = 0}^{-1} = 1$.
	\begin{lemma}
		\label{lemma:dv}
		$\alpha_e$ and $v_e$ are continuously differentiable on $\mathbb{R}_+ \setminus \{x_e,y_e\}$, and Lipschitz continuous.
		Moreover, for all $0 < x_0 < x_e$, 
		\begin{equation}
			\label{eq:dv}
			v_e'(x_0) = 
			\begin{dcases}
				\textstyle \mathbb{E}\left[ \sum_{\ell=0}^{\tilde m-1} d_e(\tilde x_{\ell})\prod_{r = 0}^{\ell-1}g_e(\tilde x_r) +  b'(\tilde x_{\tilde m})\prod_{\ell = 0}^{\tilde m-1}g_e(\tilde x_\ell) \right] & \text{if $x_0 > y_e$} \\
				\textstyle \mathbb{E}\left[ \sum_{\ell = 0}^{\tilde r-1} \left(\frac{\lambda n}{1+\lambda n}\right)^\ell\left[b'(x_\ell) - c_2(1,x_\ell)\right] + \left(\frac{\lambda n}{1+\lambda n}\right)^{\tilde r} v'_e(\tilde x_{\tilde r}) \right] & \text{if $x_0 < y_e$,}
			\end{dcases}
		\end{equation}
		where $x_\ell = x_0 + z_1 + \dots + z_\ell$ for all $\ell \ge 1$, $(z_\ell)_{\ell = 1}^\infty$ are independent draws from $F$, $m = \min\{\ell \ge 1 : x_\ell \ge x_e\}$ and $r = \min\{\ell \ge 1: x_\ell \ge y_e\}$.
		Finally, $\mathbb{E}_F[v_e(x+\tilde z)]$ is continuously differentiable in $x$ with derivative $\mathbb{E}_F[v'_e(x+\tilde z)]$.
	\end{lemma}
	\subsection{Proof of \Cref{proposition:equilibrium:general}}
	\label{proof:equilibrium:general}
	The proof of \Cref{proposition:equilibrium:general} relies on several intermediate results: \Cref{lemma:equilibrium:dp,lemma:equilibrium:p,lemma:equilibrium:decr,lemma:equilibrium:a}. I state and prove these, then prove \Cref{proposition:equilibrium:general}.
	\begin{lemma}
		\label{lemma:equilibrium:dp}
		Given a Markov strategy $\alpha$ of the game with forced disclosure and $v \in V$, 
		$\alpha$ is a symmetric MPE and induces continuation payoff $v$ if and only if $\alpha = A_ev$ and $P_ev = v$.
		Moreover, given a Markov strategy $(\alpha,\delta)$ of the game with disposal and $v \in V$, 
		$(\alpha,\delta)$ is a symmetric MPE and induces continuation payoff $v$ if and only if $\alpha = A_dv$, $P_dv = v$, and $\delta(x,z) = 1 \mathrel{(0)}$ for all $x \ge 0$ and $F$-a.e.\ $z$ such that $v(x+z) > \mathrel{(<)} v(x)$.
	\end{lemma}

	\begin{proof}[Proof of \Cref{lemma:equilibrium:dp}]
		For the first part, fix $\alpha$ and $v \in V$.
		By Theorem 3.1.2 of \textcite{piunovskiy2020}, $\alpha$ is a symmetric MPE and induces continuation payoff $v$ if and only if (i) \eqref{eq:equilibrium} holds with $\hat v_\alpha$ replaced by $v$, for all $x$, and (ii) in this modified equation, the maximum is attained by $a = \alpha(x)$.
		Moreover, (i) is equivalent to $P_e v = v$ and (ii) is equivalent to $\alpha = A_ev$.

		For the second part, fix $(\alpha,\delta)$ and $v \in V$.
		By Theorem 46.18 of \textcite{davis2018}, $(\alpha,\delta)$ is a symmetric MPE and induces continuation payoff $v$ if and only if all of the following hold: (iv) \eqref{eq:disposal:equilibrium} holds with $\hat v_\pi$ replaced by $v$, for all $x$, and (v) in this modified equation, the maximum is attained by $(a,D) = (\alpha(x),\{z \in Z: \delta(x,z) = 1\})$, and (vi) $\delta(x,z) = 1 \mathrel{(0)}$ for all $x \ge 0$ and $F$-a.e.\ $z$ such that $v(x+z) > \mathrel{(<)} v(x)$.
		Moreover, (iv) is equivalent to $P_dv = v$ and, assuming (vi) holds, (v) is equivalent to $\alpha = A_dv$.
	\end{proof}
	%
		%
	%
	\begin{lemma}
		\label{lemma:equilibrium:a}
		$\Gamma(a,x,L_kv(x)) \le P_kv(x)$ for each $k \in \{d,e\}$ and $0 \le a \le A_kv(x)$.
	\end{lemma}
	\begin{proof}[Proof of \Cref{lemma:equilibrium:a}]
	  Fix $k$ and let 
	  $\phi: [0,1]\rightarrow \mathbb R$ be given by 
	  \begin{equation*}
	    \phi(\hat a) = \frac{b(x)-c(\hat a,x) + \lambda [\hat a + (n-1)A_kv(x)]L_kv(x) }{1 + \lambda [\hat a + (n-1)A_kv(x)]}.
	  \end{equation*}
	  Note that $\phi$ is differentiable and $\phi'(\hat a)$ has the same sign as 
	  \begin{equation}
			\label{sym_h0_bis}
			L_kv(x)-[b(x)-c(\hat a,x)] - \left[\frac 1 \lambda +\hat a + (n-1)A_kv(x)\right]c_1(\hat a,x).
	  \end{equation}
	  Then, $\phi$ is quasi-concave (since $c(\hat a,x)$ is increasing and convex in $\hat a$), and $\phi'(A_kv(x))$ has the same sign as $\gamma(A_kv(x),x,L_kv(x))$ (where $\gamma$ was defined in \eqref{eq:gamma}), so that $\phi$ is maximised at $A_kv(x)$.
		Moreover, $\phi(A_kv(x)) = P_kv(x)$, so that $P_kv(x) \ge \phi(a)$. Hence, it suffices to show that $\phi(a) \ge \Gamma(a,x,L_kv(x))$.
	  
	  Since $\phi$ is quasi-concave and maximised at $A_kv(x)$, and $a \le A_kv(x)$, we have $\phi'(a) \ge 0$. 
	  Then, \eqref{sym_h0_bis} implies that $b(x)-c(a,x) \le L_kv(x)$, so that
	  \begin{align*}
	    \phi(a) &= \frac{b(x)-c(a,x)}{1+ \lambda [a + (n-1)A_kv(x)]} + \left[1-\frac{1}{1+ \lambda [a + (n-1)A_kv(x)]}\right]L_kv(x)
			\\& \ge \frac{b(x)-c(a,x)}{1+ \lambda na} + \left[1-\frac{1}{1+ \lambda na}\right]L_kv(x)
			= \Gamma\left(a,x,L_kv(x)\right),
	  \end{align*}
	  where the inequality holds since $a \le A_kv(x)$.
	\end{proof}
	Endow $V$ with the pointwise order and the topology of pointwise convergence.
	Recall that, for each $x \ge 0$, $\alpha_*(x)$ is the smallest maximiser in \eqref{eq:benchmark}.
	\begin{lemma}
		\label{lemma:equilibrium:p}
		For each $k \in \{d,e\}$, $P_k$ maps $V$ to itself and is increasing and continuous on $V$. Moreover, $A_kv \le \alpha_*$ for all $v \in V$.
	\end{lemma}
	\begin{proof}[Proof of \Cref{lemma:equilibrium:p}]
		Fix $k$. To show that $P_k$ maps $V$ to itself, fix $v \in V$ and note that $P_kv$ is measurable since $v$ is.
		Moreover, 
		\begin{equation}
			\label{eq:equilibrium_effort_bounded}
			L_kv \le L_k v_* = L_ev_*,
		\end{equation}
		as $v \le v_*$ and $v_*$ is increasing (\Cref{proposition:benchmark}).
		Then, as $\Gamma(a,x,\ell)$ increases in $\ell$, 
		\begin{equation*}
			P_kv(x) \le \Gamma\left(A_kv(x),x,L_ev_*(x)\right) \le \max_{a \in [0,1]} \Gamma (a,x,L_ev_*(x)) = v_*(x).
		\end{equation*}
		Note also that 
		$P_kv(x) \ge \Gamma(0,x,L_kv(x)) = b(x)$ by \Cref{lemma:equilibrium:a}, 
		so that $P_kv \in V$.

		To show that $P_k$ is increasing, fix $v \le w$ in $V$ and note that $L_kw(x) \le L_kv(x)$.
		Then, $P_kw(x) \le \Gamma(A_kw(x),x,L_kv(x))$ as $\Gamma(a,x,\ell)$ is increasing in $\ell$, and $A_kw(x) \le A_kv(x)$ as $p(x,\ell)$ is increasing in $\ell$ (since $\gamma(a,x,\ell)$ is).
		Hence, $\Gamma(A_kw(x),x,L_kv(x)) \le P_kv(x)$ by \Cref{lemma:equilibrium:a}, so that $P_kw(x) \le P_kv(x)$.

		To show that $P_k$ is continuous, note that $v \mapsto L_kv(x)$ is a continuous map $V \to \mathbb{R}_+$ for any $x$, by the dominated convergence theorem, since $b$ is positive and concave, $v_*-b$ is bounded (\Cref{lemma:benchmark} in \Cref{proof:benchmark}), and $F$ has finite mean.
		Moreover, $p$ and $\Gamma$ are continuous, so that $P_k$ is continuous on $V$.

		For the last part, recall from the proof of \Cref{proposition:benchmark} in \Cref{proof:benchmark} that $\gamma^*(a,x)$ is decreasing in $a$, and $\gamma^*(\alpha_*(x),x) \le 0$ whenever $\alpha_*(x) < 1$.
		Then no $v \in V$ and $x$ can satisfy $A_kv(x) > \alpha_*(x)$, or else $\gamma^*(A_kv(x),x) > \gamma(A_kv(x),x,L_kv(x)) \ge 0$, where the first inequality follows from \eqref{eq:equilibrium_effort_bounded} with $v = v_k$
		since $c_1(A_kv(x),x) > 0$ (as $A_kv(x) > \alpha_*(x)$), and the second holds since $A_kv(x) > 0$.
	\end{proof}
	\begin{lemma}
		\label{lemma:equilibrium:decr}
		For all $v \in V$ such that $v-b$ is decreasing, and $k \in \{d,e\}$, $A_kv$ and $P_kv-b$ are decreasing.
	\end{lemma}
	\begin{proof}[Proof of \Cref{lemma:equilibrium:decr}]
		Fix $k$ and $v$.
		To show that $A_kv$ is decreasing, note that $L_kv - b$ is decreasing since  $b$ is concave and $v-b$ is decreasing.
		Then, $\gamma(a,x,L_kv(x))$ is decreasing in $x$, as $c_1(a,x)$ and $c_{11}(a,x)$ are increasing in $x$. Hence, $A_kv$ is decreasing.

		It remains to show that $P_kv(x_2)-P_kv(x_1) \le b(x_2)-b(x_1)$ for all $0 \le x_1 \le x_2$.
	  Fix $x_1$ and $x_2$ and note that
	  \begin{align*}
			%
			P_kv(x_2)-P_kv(x_1) \le \Gamma\left(A_kv(x_2),x_2,L_kv(x_2)\right) -\Gamma\left(A_kv(x_2),x_1,L_kv(x_1)\right)
	    \\\pushright{\le \frac{b(x_2)-b(x_1)+\lambda nA_kv(x_2)\left[L_kv(x_2)-L_kv(x_1)\right]}{1+\lambda nA_kv(x_2)}
	    \le b(x_2)-b(x_1)},
	  \end{align*}
	  where the first inequality follows from \Cref{lemma:equilibrium:a} as $A_kv(x_2) \le A_kv(x_1)$,
		the second holds as $c(a,x)$ is increasing in $x$,
		and the third since $L_kv - b$ is decreasing.
	\end{proof}
	\begin{proof}[Proof of \Cref{proposition:equilibrium:general}]
		I first show that, for each $k \in \{d,e\}$, $v_k$ is well defined and is the unique fixed point of $P_k$ in $V$.
		This delivers the first part of \Cref{proposition:equilibrium:general} and guarantees that $\alpha_k = A_kv_k$ and $v_k = P_kv_k$ for each $k \in \{d,e\}$, by \Cref{lemma:equilibrium:dp}.
		Fix $k$.
	  Note that $v_k$ and $\bar v_k = \lim_{m \to \infty} (P_k)^m v_*$ are well defined, lie in $V$, and are fixed points of $P_k$, since $P_k$ is increasing and continuous (by \Cref{lemma:equilibrium:p}) and $V$ is closed.
	  Moreover, they are the smallest and largest fixed points of $P_k$ in $V$, since $P_k$ is increasing and $b \le v \le v_*$ for all $v \in V$.
	  Then, it suffices to show that $v_k = \bar v_k$.
	  Since $b \le v_k \le \bar v_k \le v_*$ and $v_* = b$ on $[x_*,\infty)$ (\Cref{proposition:benchmark}), it is enough to show that
		\begin{equation}
			\label{eq:ranked}
			\bar v_k(x)-v_k(x) \le L_k \bar v_k(x) - L_k v_k(x)
		\end{equation}
		for all $x \ge 0$.
		If $k = d$, \eqref{eq:ranked} suffices because, given any $\epsilon > 0$, there exists $\delta > 0$ such that $\bar v_k(x) \le b(x) + \epsilon$ for all $x$ satisfying $\Pr_F(\bar v_k(x+\tilde z) > \bar v_k(x)) < \delta$.%
		\footnote{
			\label{footnote:ranked}
			To understand why \eqref{eq:ranked} suffices if $k = d$, assume by contradiction that $\Delta = \sup (\bar v_k - v_k) > 0$ and consider a sequence $x_m$ with limit $\inf\{\hat x \in \mathbb{R}_+ : \sup\{\bar v_k(y)-v_k(y) : y \ge \hat x\} < \Delta\}$ such that $\bar v_k(x_m)-v_k(x_m) \rightarrow \Delta$, let $E_m = \{z \in \mathbb{R}_+ : \bar v_k(x_m+z) > \bar v_k(x_m)\}$ and $p = \lim\inf_m \Pr_F(E_m)$.
			Note that $p > 0$ since $b \le v_k$, so that \eqref{eq:ranked} yields $\Delta \le \lim \sup_m L_k \bar v_k(x_m) - L_k v_k(x_m) \le p \lim \sup_m \mathbb{E}_F[\bar v_k(x_m+\tilde z) - v_k(x_m+\tilde z)|E_m] + (1-p)\Delta < \Delta$, which is absurd. 
			}
		Let $\bar \alpha_k = A_k\bar v_k$ and note that $\bar \alpha_k \ge \alpha_k$, since $\bar v_k \ge v_k$ and $p(x,l)$ is increasing in $l$.
		Fix $x$ and suppose first that $\bar \alpha_k(x) = 0$.
		Then, $\alpha_k(x) = 0$ since $\bar \alpha_k \ge \alpha_k$,
		so that $\bar v_k(x) = b(x) = v_k(x)$. Therefore, \eqref{eq:ranked} holds since $\bar v_k \ge v_k$.
		If $\alpha_k(x) = 1$, then $\bar \alpha_k(x) = 1$, so that
		\begin{equation*}
			\left[L_k\bar v_k(x) - \bar v_k(x)\right] - \left[L_kv_k(x) - v_k(x)\right] = \frac {\bar v_k(x)-v_k(x)}{\lambda n} \ge 0,
		\end{equation*}
		where the equality follows from \eqref{eq:equilibrium} and \eqref{eq:disposal:equilibrium}. 
		Thus, \eqref{eq:ranked} holds.
		Finally, if $\bar \alpha_k(x) > 0$ and $\alpha_k(x) < 1$, then
		\begin{equation*}
			L_k \bar v_k(x) - \bar v_k(x) \ge \frac {c\left(\bar \alpha_k(x),x\right)} \lambda  \ge \frac {c\left(\alpha_k(x),x\right)} \lambda  \ge L_k v_k(x) - v_k(x),
		\end{equation*}
		where the first inequality follows from \eqref{eq:equilibrium} and \eqref{eq:disposal:equilibrium}, since $\bar \alpha_k(x) > 0$, the second inequality holds since $\bar\alpha_k(x) \ge \alpha_k(x)$, and the third follows from \eqref{eq:equilibrium} and \eqref{eq:disposal:equilibrium}, since $\alpha_k(x) < 1$. 
		Then, \eqref{eq:ranked} holds.

		Note that $\alpha_k$ and $v_k-b$ are decreasing since $(A_k)^mb$ and $(P_k)^mb-b$ are decreasing for each $m \ge 0$, by \Cref{lemma:equilibrium:decr}.
		To prove that $v_k$ is continuous, note that it has bounded variation since $v_k \ge b$ and $v_k-b$ is decreasing.
		Then, we may define $\bar v,\underline v : \mathbb{R}_+ \rightarrow \mathbb{R}$ by 
	  \begin{equation*}
	      \bar v(x) = \lim_{y \downarrow x} v_k(y) \quad \& \quad
	      \underline v(x) =
	      \begin{cases}
	      v_k(0) & \text{if $x = 0$}\\
	      \lim_{y \uparrow x}v_k(y) & \text{if $x > 0$.}
	      \end{cases}
	  \end{equation*}
	  It is easy to see that $\bar v$ and $\underline v$ are fixed points of $P_k$, so that $\bar v = v_k = \underline v$.
	  Then $v_k$ is continuous.

	  As noted above, $\alpha_k$ is decreasing, and $\alpha_k \le \alpha_*$ by \Cref{lemma:equilibrium:p}.
	  To prove that $\alpha_k$ is continuous, note that $L_kv_k$ is continuous by the dominated convergence theorem, since $v_k$ is continuous, $b \le v_k \le v_*$, and $v_*-b$ is bounded (\Cref{lemma:benchmark} in \Cref{proof:benchmark}).
	  Then, $\alpha_k$ is continuous since $p$ is.

	  The fact that \eqref{eq:equilibrium_cutoff} admits a unique solution $x_e > 0$ follows from \Cref{proposition:benchmark}, since the proof of the latter is valid for $n = 1$. Then, $x_e < x_*$ since $c_1(0,x_*) > c_1(0,0) = 0$ where the second inequality holds since $x_* > 0$ and $c_{12} >0$.
		To show that $\alpha_k$ vanishes on $[x_e,\infty)$, note that
		\begin{equation*}
			\lambda[L_kv_k(x)-v_k(x)] \le \lambda[L_kb(x)-b(x)] = \lambda\{\mathbb{E}_F[b(x+\tilde z)]-b(x)\} < c_1(0,x)
		\end{equation*}
		for any $x > x_e$, 
		where the first inequality holds since $v_k-b$ is decreasing, the second since $b$ is increasing, and the last since $b$ is concave and is $c_1(0,x)$ strictly increasing in $x$.
		Then, $\alpha_k(x) = 0$ by \eqref{eq:equilibrium} and \eqref{eq:disposal:equilibrium}, so that $\alpha_k$ vanishes on $[x_e,\infty)$, since $\alpha_k$ is continuous.
	\end{proof}
	\subsection{Proof of \Cref{lemma:dv}}
	\label{proof:dv}
	I state and prove an auxiliary result (\Cref{lemma:dp} below), then prove \Cref{lemma:dv}. 
	\begin{lemma}
		\label{lemma:dp}
		Let $v \in V$ be Lipschitz continuous and $v' : \mathbb{R}_+ \to \mathbb{R}$ a.e.-equal to the derivative of $v$.
		Then, $\mathbb{E}_F[v(x+\tilde z)]$, $A_ev(x)$ and $P_ev(x)$ are Lipschitz continuous in $x$, $\mathbb{E}_F[v'(x+\tilde z)]$ is a.e.-equal to the derivative of $\mathbb{E}_F[v(x+\tilde z)]$ and, for a.e.\ $x$, 
		\begin{equation}
			\label{eq:dv_rec}
			(P_ev)'(x) =
			\begin{dcases} 
				d(A_ev(x),x) + g(A_ev(x),x)\mathbb{E}_F[v'(x + \tilde z)] & \text{if $A_ev(x) \in (0,1)$}\\
				\frac{b'(x)-c_2(1,x) + \lambda n\mathbb{E}_F[v'(x+\tilde z)]}{1+\lambda n}  & \text{if $A_ev(x) = 1$.}
			\end{dcases}
		\end{equation} 
	\end{lemma}
	\begin{proof}[Proof of \Cref{lemma:dp}]
		Note that $\mathbb{E}_F[v(x+\tilde z)]-\mathbb{E}_F[v(\tilde z)] = \mathbb{E}_F\left[\int^x_0 v'(y+\tilde z)\text{d}y\right] = \int_0^x \mathbb{E}_F[v'(y+\tilde z)] \text{d}y$
		for all $x$, where the first equality holds since $v$ is Lipschitz and the second since $v'$ is bounded and measurable. Then, $\mathbb{E}_F[v(x+\tilde z)]$ is Lipschitz in $x$ with derivative a.e.-equal to $\mathbb{E}_F[v'(x+\tilde z)]$.

		To prove that $A_ev$ is Lipschitz, note that there is $\epsilon > 0$ such that $c_1(A_ew(x),x) > \epsilon$ for all $x \ge 0$ and $w \in V$ satisfying $p(x,L_ew(x)) < 1$, since $c_1(a,x)$ is increasing in $a$ and strictly increasing in $x$, and $c_1(A_ew(0),0) \ge c_1(A_eb(0),0) \ge \lambda \{\mathbb{E}_F[b(\tilde z)]-b(0)\} > 0$ for any $w\in V$ with $p(0,L_ew(0)) < 1$.
		Then, $p$ is continuously differentiable with bounded derivative on $\{(x,\ell) \in [0,x_*] \times (b(0),\infty) : p(x,\ell) < 1\}$, by the inverse function theorem, since $b$ and $c$ have locally-bounded first and second derivatives.
		Moreover, $A_ev(x) \le \alpha_*(x) = 0$ for $x \ge x_*$, where the inequality follows from \Cref{lemma:equilibrium:p} and the equality from \Cref{proposition:benchmark}.
		Then, $A_ev$ is Lipschitz since $L_ev$ is Lipschitz and $p$ is continuous.

		To prove that $P_ev$ is Lipschitz, note that it is Lipschitz on $[x_*,\infty)$ since $A_ev$ vanishes on this interval and $b$ is Lipschitz.
		Moreover, $P_ev$ is Lipschitz on $[0,x_*]$ since $\Gamma$ has locally-bounded derivatives and $v(x)$, $A_ev(x)$ and $\mathbb{E}[v_e(x+\tilde z)]$ are Lipschitz in $x$.
		Hence, $P_ev$ is Lipschitz.

		To establish \eqref{eq:dv_rec} note that
		\begin{equation*}
			P_ev(x) = \max_{a \in [0,1]} b(x)-c(a,x) + \lambda[a + (n-1)A_ev(x)]\{\mathbb{E}_F[v(x+\tilde z)]-P_ev(x)\}
		\end{equation*}
		and the maximum is attained by $a = A_ev(x)$, so that $\lambda\{\mathbb{E}_F[v(x+\tilde z)]-P_ev(x)\} = c_1(A_ev(x),x)$
		for any $x$ such that $0 < A_ev(x) < 1$.
		Since $\mathbb{E}_F[v(x+\tilde z)]$, $A_ev(x)$ and $P_ev(x)$ are Lipschitz in $x$, for a.e.\ $x$, 
		\begin{align*}
			&(P_ev)'(x) = b'(x)-c_2(A_ev(x),x) + \lambda n A_ev(x)\{\mathbb{E}_F[v'(x+\tilde z)]-(P_ev)'(x)\}
			\\& \pushright{+ \; \lambda(n-1)(A_ev)'(x)\{\mathbb{E}_F[v(x+\tilde z)]-P_ev(x)\} \text{, while}}
			\\&c_{11}(A_ev(x),x)(A_ev)'(x) = \lambda\{\mathbb{E}_F[v'(x+\tilde z)]-(P_ev)'(x)\} - c_{12}(A_ev(x),x)
		\end{align*}
		for a.e.\ $x$ such that $0 < A_ev(x) < 1$.
		Combining the last three equations yields \eqref{eq:dv_rec}, since $(A_ev)'(x) = 0$ for a.e.\ $x$ such that $A_ev(x) = 1$.
	\end{proof}
	\begin{proof}[Proof of \Cref{lemma:dv}]
		I begin by showing that $v_e$ is Lipschitz.
		Recall that $v_e = \lim_{m \to \infty} w_m$ where $w_m = (P_e)^m b$ is well defined and lies in $V$ for each $m \in \{0,1,\dots\}$, by \Cref{proposition:equilibrium:general}.
		Moreover, $w_m$ is Lipschitz for each $m$ by \Cref{lemma:dp}, since $b$ is.
		Then, it suffices to exhibit $\ell > 0$ such that $|w_m'(x)| \le \ell$ for all $m \in \{0,1,\dots\}$ and a.e.\ $x \le x_*$, 
		as $v_e$ is Lipschitz on $[x_*,\infty)$ by \Cref{proposition:equilibrium:general}.
		
		In light of \eqref{eq:dv_rec} and because $b$ is Lipschitz and $c$ has locally-bounded second derivatives, there exists $\beta > x_*$ such that $|b'(x)| < \beta-x_*$ and $|w_{m+1}'|(x) \le |\mathbb{E}_F[w_m'(x+\tilde z)]| + \beta$ for all $m \in \{0,1,\dots\}$ and a.e.\ $x \le x_*$.
		Since $\beta-x_* > b'(x_*)$ and $b$ is concave, there also exists $0 < \epsilon < \beta$ such that $\mathbb{E}_F[\max\{\beta-(x+\tilde z),b'(x+\tilde z)\}] < \beta-x - \epsilon$ for all $x \le x_*$.
		It is enough to show that $|w_m'(x)| \le \frac \beta \epsilon(\beta-x)$ for all $m \in \{0,1,\dots\}$ and a.e.\ $x \le x_*$, as we may then pick $\ell = \beta^2 /\epsilon$.
		Proceed by induction on $m$. The base case $m = 0$ is immediate.
		For the induction step, suppose that the relation holds for $m$, note that $|w'_m(x)| \le \frac \beta \epsilon\max\{\beta-x,b'(x)\}$ for a.e.\ $x \ge 0$, so that  
		\begin{align*}
			\left|w'_{m+1}(x)\right| 
			&\le |\mathbb{E}_F[w'_m(x+\tilde z)]| + \beta 
			\\&\le \frac \beta \epsilon\mathbb{E}_F[\max\{\beta-(x+\tilde z),b'(x+\tilde z)\}] + \beta \le \frac \beta \epsilon\left(\beta-x\right)
		\end{align*}
		for a.e. $x\le x_*$.

		Next, I argue that \eqref{eq:dv} holds for a.e.\ $y_e < x_0 < x_e$, as well as for a.e.\ $0 < x_0 < x_e$.
		For the former note that, by \Cref{lemma:dp} and as $P_ev_e = v_e$ (\Cref{proposition:equilibrium:general}) and $v_e$ is Lipschitz, given a bounded $v_e' : \mathbb{R}_+\to\mathbb{R}$ a.e.-equal to the derivative of $v_e$,
		\begin{equation*}
			v_e'(x_0) = \textstyle \mathbb{E}\left[ \sum_{\ell=0}^{\tilde m_L-1} d_e(\tilde x_{\ell})\prod_{r = 0}^{\ell-1}g_e(\tilde x_r) +  v_e'(\tilde x_{\tilde m_L})\prod_{\ell = 0}^{\tilde m-1}g_e(\tilde x_\ell) \right]
		\end{equation*}
	  for all $L \in \{1,2,\dots\}$ and a.e.\ $y_e < x_0 < x_e$, 
		where $m_L = \min(\{\ell \ge 1: x_\ell \ge x_e\} \cup \{L\})$.
		Taking the limit as $L \to \infty$ yields \eqref{eq:dv}, by the bounded convergence theorem, since $d_e$ and $g_e$ are bounded on $(0,x_e]$ and $v_e$ is Lipschitz. 
		For the latter note that, by \Cref{lemma:dp},
		\begin{equation*}
			v_e'(x_0) = \textstyle \mathbb{E}\left[ \sum_{\ell = 0}^{\tilde r_L-1} \left(\frac{\lambda n}{1+\lambda n}\right)^\ell\left[b'(x_\ell) - c_2(1,x_\ell)\right] + \left(\frac{\lambda n}{1+\lambda n}\right)^{\tilde r_L} v'_e\left(\tilde x_{\tilde r_L}\right) \right]
		\end{equation*}
		holds for all $L \in \{1,2,\dots\}$ and a.e.\ $0 < x_0 < y_e$, where $r_L = \min(\{\ell \ge 1: x_\ell \ge y_e\} \cup \{L\})$. 
		Taking the limit as $L \to \infty$ yields \eqref{eq:dv}, by the bounded convergence theorem, since $v_e$ is Lipschitz.

		Note that it is continuously differentiable on $(0,y_e) \cup (y_e,x_e)$ and \eqref{eq:dv} holds for all $0 < x_0 < x_e$, since $v_e$ is Lipschitz, \eqref{eq:dv} holds for a.e.\ $0 < x_0 < x_e$, $d_e$ and $g_e$ are continuous (as $\alpha_e$ is, by \Cref{proposition:equilibrium:general}), and $b'$, $c_2$, and $F$ are continuous as well. 
		Then, $v_e$ is continuously differentiable on $\mathbb{R}_+ \setminus \{x_e,y_e\}$ since $v_e = b$ on $[x_e,\infty)$, by \Cref{proposition:equilibrium:general}.
		Moreover, $\mathbb{E}_F[v_e(x+\tilde z)]$ is continuously differentiable in $x$ with derivative $\mathbb{E}_F[v'_e(x+\tilde z)]$, by \Cref{lemma:dp}, since $F$ is continuous.
		Therefore, $\alpha_e$ is continuously differentiable on $(0,x_e)$, by the inverse function theorem, since $c_1(\alpha_e(x),x) = \lambda \{\mathbb{E}_F[v_e(x+\tilde z)]-v_e(x)\}$ on this interval (by \Cref{proposition:equilibrium:general}).
		Hence, $\alpha_e$ is Lipschitz since it is continuous and it is constant on $(0,y_e)$ as well as on $[x_e,\infty)$, by \Cref{proposition:equilibrium:general}.
	\end{proof}
	\section{Proofs of \Cref{proposition:detrimental_n,proposition:detrimental_longrun,proposition:cost,corollary:detrimental}}
	\label{proof:detrimental}
	I prove \Cref{proposition:detrimental_n,proposition:detrimental_longrun,corollary:detrimental,proposition:cost}.
	\begin{proof}[Proof of \Cref{proposition:detrimental_n}]
		Let $\alpha^n_e$ and $v^n_e$ be effort and continuation payoff, respectively, in the symmetric MPE with $n$ agents, and set $y^n_e = \inf\{x \ge 0: \alpha^n_e(x) < 1\}$.
		Suppose that $v^n_e$ is increasing for all $n$, and seek a contradiction.
		Because $v^n_e-b$ is decreasing (\Cref{proposition:equilibrium:general}) and $b$ concave with $b'(0) < \infty$, $v^n_e$ is $b'(0)$-Lipschitz. 
		Then, $v^n_e$ converges pointwise to a $b'(0)$-Lipschitz map $v : \mathbb{R}_+ \to \mathbb{R}$, by the Arzela-Ascoli theorem. Moreover, $v$ is increasing and $v-b$ is decreasing.
		
		Note that $x_e$ does not vary with $n$.
		Then, $\lim_{n \to \infty}\alpha_e(x) = 0$ for all $x > 0$, for if $\lim_{n \to \infty}\alpha^n_e(x) > 0$ as $n \to \infty$ for some $x > 0$ and along some sequence, then 
		\begin{equation*}
			\mathbb{E}_F[v^n_e(x+\tilde z)]-v^n_e(x) = \frac{\mathbb{E}_F[v^n_e(x+\tilde z)] - [b(x)-c(\alpha^n_e(x),x)]}{1 + \lambda n\alpha^n_e(x)} 
			\to 0
		\end{equation*}
		along this sequence, since $v^n_e$ is $b'(0)$-Lipschitz with $v^n_e(x_e) = b(x_e)$, and this contradicts the fact that $a = \alpha^n_e(x)$ attains the maximum in \eqref{eq:equilibrium} for all $n$, since $c_1(0,x) > 0$.
		Then, $\lim_{n \to \infty} y^n_e = 0$. Hence, letting $n \to \infty$ in the first-order condition derived from \eqref{eq:equilibrium} 
		yields $\mathbb{E}_F[v(x+\tilde z)] - v(x) = \textstyle \frac 1 \lambda c_1(0,x)$
		for all $0 < x < x_e$, by dominated convergence, and letting $x \to 0$ yields $v(0) = \mathbb{E}_F[v(\tilde z)]$. 
		Note that $v^n_e = b$ on $[x_e,\infty)$ for each $n$, by \Cref{theorem:equilibrium}, so that $v = b$ on $[x_e,\infty)$.
		Then, $\mathrm{supp}(F) = [0,\hat x]$ for some $\hat x \le x_e$ and $v$ is constant on $[0,\hat x]$, since $b'(x_e) > 0$, $v$ is increasing and continuous, and $\mathrm{supp}(F)$ is an interval containing $0$.
		Hence, 
		\begin{equation*}
			\frac {c_1(0,x)}\lambda  = \mathbb{E}_F[v(x+\tilde z)]-v(x) \le b'(0)\mathbb{E}_F[\max\{x+z-\hat x,0\}] \le x b'(0)[1-F(\hat x-x)]
		\end{equation*}
		for any $0 < x < \hat x$, where the first inequality holds since $v$ is $b'(0)$-Lipschitz, increasing, and constant on $[0,\hat x]$, and the second holds since $F(\hat x) = 1$.
		Dividing by $x$ and letting $x \to 0$ yields $c_{12}(0,0)/\lambda \le 0$ since $F$ is continuous, which is impossible. 
	\end{proof}
	\begin{proof}[Proof of \Cref{proposition:detrimental_longrun}]
		By \Cref{lemma:dv} in \Cref{proof:equilibrium}, it suffices to show that the top expression in the right-hand side of \eqref{eq:dv} converges to the right-hand side of \eqref{eq:d_longrun} as $x_0$ tends to $x_e$ from below. 
		To this end, fix $0 < x < x_e$ and note that $d_e$ and $g_e$ are continuous on $[x,x_e]$, since $\alpha_e$ is continuous (\Cref{theorem:equilibrium}) and $c_1$ is increasing with $c_1(0,x) > 0$.
		Then, since $b'$ is bounded, the top expression in the right-hand side of \eqref{eq:dv} converges to $d_e(x_e) + \mathbb{E}_F[b'(x_e+\tilde z)]g_e(x_e)$ as $x$ tends to $x_e$ from below, by the bounded convergence theorem.
		Since $\alpha_e(x_e) = 0$ (\Cref{theorem:equilibrium}), this expression is equal to the right-hand side of \eqref{eq:d_longrun}.
	\end{proof}
	\begin{proof}[Proof of \Cref{corollary:detrimental}]
		Let $\psi : (0,\infty) \to \mathbb{R}$ be given by $\psi(\lambda') = \lambda' \mathbb{E}_{F_{\lambda'}}[b'(\hat x+\tilde z)]$.
		Note that $c_1(0,\hat x) > 0$ since $\hat x > 0$, so that $b'(\hat x) > 0$ by \eqref{eq:equilibrium_cutoff_lambda}, since $b'$ is positive and decreasing. 
		Then, $\psi(\lambda') >0$ for all $\lambda' >0$ since $b'$ is continuous and the support of $F_{\lambda'}$ includes $0$.
		Hence, $\psi$ is strictly increasing since $b'$ is decreasing and $(F_{\lambda'})_{\lambda' > 0}$ is reverse FOSD-ordered, with $\psi(\lambda') \to 0 \mathrel{(\infty)}$ as $\lambda' \to 0 \mathrel{(\infty)}$.
		Then, there exists a unique $\lambda^* \in \mathbb{R}_+$ such that $b'(\hat x)c_{11}(0,\hat x) + (n-1)c_1(0,\hat x)[\psi(\lambda') - c_{12}(0,\hat x)] > \mathrel{(<)} 0$
		for all $\lambda' \in \mathbb{R}$ satisfying $\lambda' > \mathrel{(<)} \lambda^*$, with $\lambda^* > 0$ if \eqref{eq:detrimental_lambda_n} holds.
		
		Note that, if $F = F_{\lambda}$, then $x_e = \hat x$ by \Cref{theorem:equilibrium}, so that 
		\begin{equation*}
			v'_e(x^-_e) = \frac{b'(\hat x)c_{11}(0,\hat x) + (n-1)c_1(0,\hat x)[\psi(\lambda) - c_{12}(0,\hat x)]}{c_{11}(0,\hat x) + \lambda (n-1)c_1(0,\hat x)},
		\end{equation*}
		by \Cref{proposition:detrimental_longrun}.
		Result follows.
	\end{proof}
	\begin{proof}[Proof of \Cref{proposition:cost}]
		I first argue that the proof of \Cref{theorem:equilibrium} remains valid with minimal changes.
		To this end, note that the results in the following list are established without reference to any result that is not in the list: 
		\Cref{proposition:benchmark,theorem:equilibrium,lemma:benchmark,proposition:equilibrium:general,lemma:dv,lemma:equilibrium:dp,lemma:equilibrium:p,lemma:equilibrium:decr,lemma:equilibrium:a,lemma:dp}.
		Moreover, for the purposes of establishing all of the listed results, the assumptions that $c_{12}(a,x)$ is strictly instead of weakly positive and that $c_1(0,0) =0$ are used precisely four times: 
			(i) to ensure that $\psi$ is strictly decreasing with $\psi(0) = 0$ in the proof of \Cref{proposition:benchmark} (in \Cref{proof:benchmark}),
			(ii) to ensure that $c_1(a,x) > 0$ for all $0 \le a \le 1$ and $x > 0$ when defining $d$ and $g$ in \Cref{proof:equilibrium},
			(iii) to ensure that $c_1(0,x_*) > 0$ and that $\lambda \{\mathbb{E}_F[b(x+\tilde z)]-b(x)\} < c_1(0,x)$ for all $x < x_e$, in the last paragraph of the proof of \Cref{proposition:equilibrium:general}, and 
			(iv) to ensure that there exists $\epsilon > 0$ such that $c_1(A_ew(x),x) > \epsilon$ for all $x \ge 0$ and $w \in V$ satisfying $p(x,L_ew(x)) < 1$, in the proof of \Cref{lemma:dp}.
		Finally, it is clear that all four properties are preserved under the hypotheses of \Cref{proposition:cost}.

		It remains to prove that $v_e$ is increasing. This follows from \Cref{lemma:dv}.
	\end{proof}
	\section{Proofs of \Cref{lemma:linear,proposition:linear_lambda,proposition:linear_M,proposition:linear_cs}}
	\label{proof:linear}
	I prove \Cref{lemma:linear,}, then state an prove an auxiliary result (\Cref{lemma:y}), then prove \Cref{proposition:linear_lambda,proposition:linear_M}.
	I then state and prove another intermediary result (\Cref{lemma:equilibrium:ge}), then prove \Cref{proposition:linear_cs}. (I will use of \Cref{lemma:equilibrium:ge} to prove \Cref{theorem:disposal:equilibrium} as well).
	\begin{proof}[Proof of \Cref{lemma:linear}]
		Immediate from \Cref{theorem:equilibrium} and \Cref{lemma:dv} in \Cref{proof:equilibrium}.
	\end{proof}
	\begin{lemma}
		\label{lemma:y}
		If \eqref{eq:linear} holds, then $v_e(y_e) = n y_e$ and $y_e = [\lambda(n-1)]^{-1} \int_{0}^{\lambda\mu-y_e} M_F(\Delta)\text{d}\Delta$.
	\end{lemma}
	\begin{proof}[Proof of \Cref{lemma:y}]
		Suppose that \eqref{eq:linear} holds.
		To show that $v_e(y_e) = n y_e$, note that $v_e(x) = n\alpha(x)x$ for any $y_e < x < \lambda\mu$, by \eqref{eq:equilibrium}, as $0 < \alpha(x) < 1$.
		Letting $x$ converge to $y_e$ from above yields $y_e > 0$, since $v_e$ is continuous (\Cref{theorem:equilibrium}) and satisfies $v_e(0) > 0$ (as $b(0) = 0$ and $b'(0) > 0 = c_1(0,0)$). 
		Then, $v_e(y_e) = ny_e$ since $\alpha_e$ is continuous as well (\Cref{theorem:equilibrium}).
		
		To complete the proof, note that $v_e(y_e) = y_e + \frac 1 \lambda \int_{0}^{\lambda\mu-y_e} M_F(\Delta)\text{d}\Delta$, where the equality follows by \Cref{lemma:linear} and the fundamental theorem of calculus, since $v_e$ is Lipschitz continuous with $v_e(\lambda\mu) = \lambda\mu$ (\Cref{theorem:equilibrium}).
		Result follows since $v_e(y_e) = ny_e$.
	\end{proof}
	\begin{proof}[Proof of \Cref{proposition:linear_lambda}]
		For the first part, note that $v_e$ is strictly decreasing on $[y_e,\lambda\mu]$ if $\lambda \le 1$, by \Cref{theorem:equilibrium,lemma:linear}, since the support of $F$ includes $0$.
		Then, result follows since $\lim_{n \to \infty} y_e = 0$, by \Cref{lemma:y}.

		For the second part, note that 
		\begin{equation*}
			ny_e = y_e + \frac 1 \lambda \int_{0}^{\lambda\mu-y_e} M_F(\Delta)\text{d}\Delta \ge y_e + \frac 1 {\lambda\mu} \int_{0}^{\lambda\mu-y_e} \Delta\text{d}\Delta \ge \frac {\lambda\mu}2,
		\end{equation*}
		where the equality follows from \Cref{lemma:y} and the first inequality from Corollary 2.2(iii) of \textcite{barlow1964}.
		Note also that $\limsup_{\lambda \to \infty} M_F([1-1/(2n)]\lambda\mu)/\lambda < 1$, by the elementary renewal theorem.
		Then $v_e$ is increasing on $[y_e,\infty)$ for $\lambda$ sufficiently large, by \Cref{lemma:linear,theorem:equilibrium}, since $M_F$ is increasing.
		Result follows from \Cref{lemma:dv}.
	\end{proof}
	\begin{proof}[Proof of \Cref{proposition:linear_M}]
		Set $\delta = \mu(1-1/n)$ and note that
		\begin{equation*}
			M_F\left(\delta\lambda\right) \ge \frac{\delta\lambda}{\int_0^{\delta\lambda}1-F(z) \,\mathrm{d}z} \ge \frac{\lambda
			}{\lambda-\frac{1}{\delta}\int_0^{\delta\lambda} F(z) \,\mathrm{d}z} > \lambda,
		\end{equation*}
		where the first inequality follows from Theorem 2.4(i) of \textcite{barlow1964} and the second from \eqref{eq:linear_onlyif}.
		Then, if $y_e \le \lambda\mu/n$, $v_e'(x) < 0$ for $x > y_e$ sufficiently close to $\lambda\mu/n$, by \Cref{theorem:equilibrium,lemma:linear}, since $M_F$ is continuous.
		Hence, we may assume without loss that $y_e > \lambda\mu/n$.
		Then $v_e(y_e) = n y_e > \lambda\mu = v_e(\lambda\mu)$, where the first equality follows from \Cref{lemma:y}.
		Hence $v_e$ is not increasing on $[0,\lambda\mu]$.
	\end{proof}
	Recall \Cref{proposition:equilibrium:general} and the definitions of $L_e$, $L_d$, $p$ and $\Gamma$ from \Cref{proof:equilibrium}. 
	\begin{lemma}
		\label{lemma:equilibrium:ge}
		Let $k \in \{d,e\}$, $v \in V$ and $w: \mathbb{R}_+ \to (0,\infty)$ be such that $L_kv \ge w$ and $v(x) = \Gamma(p(x,w(x)),x,w(x))$ for all $x$.
		Then $v_k(x) \ge v(x)$ and $\alpha_k(x) \ge p(x,w(x))$ for all $x$, with the first inequality strict if $L_kv(x) > w(x)$ and $p(x,w(x)) > 0$.
	\end{lemma}
	\begin{proof}[Proof of \Cref{lemma:equilibrium:ge}]
		Note that $A_kv(x) = p(x,L_kv(x)) \ge p(x,w(x))$ for all $x$, since $L_kv(x) \ge w(x)$ and $p(x,\ell)$ is increasing in $\ell$.
		Then 
		\begin{equation*}
			P_kv(x) \ge \Gamma(p(x,w(x)),x,L_kv(x)) \ge \Gamma(p(x,w(x)),x,w(x)) = v(x)
		\end{equation*}
		for all $x$, where the first inequality follows from \Cref{lemma:equilibrium:a} and the second holds since $\Gamma(a,x,\ell)$ is increasing in $\ell$.
		Moreover, the second inequality is strict if $L_kv(x) > w(x)$ and $p(x,w(x)) > 0$.
		Then, \Cref{lemma:equilibrium:p} implies that $\bar v = \lim_{m \to \infty} (P_k)^mv$ is a fixed point of $P_k$ in $V$, that $\bar v \ge v$, and that $\bar v(x) > v(x)$ for all $x$ such that $L_kv(x) > w(x)$ and $p(x,w(x)) > 0$.
		Hence, $\bar v = v_k$ and $A_k\bar v = \alpha_k$ by \Cref{lemma:equilibrium:dp,proposition:equilibrium:general}.
		Moreover, $L_k\bar v(x) \ge L_kv(x) \ge w(x)$ for all $x$, so that $\alpha_k(x) = p(x,L_k\bar v(x)) \ge p(x,w(x))$, where the inequality holds as $p(x,\ell)$ increases in $\ell$.
	\end{proof}
	\begin{proof}[Proof of \Cref{proposition:linear_cs}]
		Note that $\mathbb{E}_F[v_e(x+\tilde z)] \ge b(x) \ge 0$ for all $x \ge 0$, where the first inequality is strict if $x = 0$ and the second one if $x > 0$, since $v_e \ge b$ and \eqref{eq:linear} holds.
		Then, it suffices to show that $\mathbb{E}_{F^\dag}[v_e(x+\tilde z)] \ge \mathbb{E}_F[v_e(x+\tilde z)]$ for all $x$, as this implies that 
		$v_e^\dag(x) \ge \Gamma(p(x,\mathbb{E}_F[v_e(x+\tilde z)]),x,\mathbb{E}_F[v_e(x+\tilde z)]) = v_e(x)
		$
		and that $\alpha_e^\dag(x) \ge p(x,\mathbb{E}_F[v_e(x+\tilde z)]) = \alpha_e(x)$ for all $x$, where the inequalities follow from \Cref{lemma:equilibrium:ge} and the equalities from \Cref{proposition:equilibrium:general}.

		To this end, let $\zeta : [0,1] \to [0,\infty]$ be given by $\zeta(r) = \inf(\{z \ge 0 : F(z) \ge r\} \cup \{\infty\})$ and $\zeta^\dag$ be the analogue of $\zeta$ for $F^\dag$.
		Note that, for $0 \le r \le 1$ such that $\zeta(r) \le \lambda\mu$, $F^\dag(\zeta(r)) \ge F(\zeta(r)) \ge r$, where the first inequality holds since $F^\dag$ is more dispersed than $F$.
		Then, $\zeta^\dag(r) \le \zeta(r)$ for any such $r$, so that 
		\begin{equation}
			\label{eq:zeta}
			v_e\left(x+\zeta^\dag(r)\right)-v_e(x+\zeta(r)) \ge b\left(x+\zeta^\dag(r)\right)-b(x+\zeta(r))
		\end{equation} 
		since $v_e-b$ is decreasing (\Cref{proposition:equilibrium:general}).
		Moreover, \eqref{eq:zeta} holds if $\zeta(r) > \lambda\mu$ as well, since $v_e \ge b$ with equality on $[\lambda\mu,\infty)$.
		Hence, 
		\begin{align*}
			&\mathbb{E}_{F^\dag}[v_e(x+\tilde z)] - \mathbb{E}_F[v_e(x+\tilde z)] = \int_0^1 v_e\left(x+\zeta^\dag(r)\right)-v_e(x+\zeta(r))\text{d}r 
			\\&\hspace{3cm}\ge \int_0^1 b\left(x+\zeta^\dag(r)\right)-b(x+\zeta(r))\text{d}r = \mathbb{E}_{F^\dag}(\tilde z) - \mathbb{E}_F(\tilde z) = 0,
		\end{align*}
		where the second equality follows from \eqref{eq:linear}, and the last holds since $F^\dag$ is more dispersed than $F$.
	\end{proof}
	%
	%
	%
	%
	\section{Proof of \Cref{proposition:sse,proposition:ppe}}
	\label{proof:ppe}
	I first state an auxiliary result (\Cref{proposition:ppe:dp}), then state and prove a strengthening of \Cref{proposition:sse} (\Cref{proposition:sse_general}), then prove \Cref{proposition:ppe}.
	I will use \Cref{proposition:sse_general} to prove \Cref{theorem:disposal:equilibrium} as well.

	Recall from \Cref{disposal:model} that I refer to the baseline model as the game \emph{with forced disclosure,} and to the model introduced in \Cref{disposal:model} as the game \emph{with disposal}.
	\Cref{proposition:ppe:dp} relies on several definitions.
	Given a strategy profile $\xi = (\sigma^i,\chi^i)_{i = 1}^n$ of the game with forced disclosure, a public history $h$, an agent $i$, and $t > t_h$, let 
	\begin{equation}
		\label{eq:payoff_t}
		v^i_{\xi,h}(t) = \mathbb E\left(\sum_{\ell = 0}^{\tilde m} \int_{\tilde t_\ell}^{\tilde t_{\ell+1}} e^{t_h-s} \left[b\left(x_{\tilde h^\ell}\right) - c\left(\sigma^i_{\tilde h^\ell}(s),x_{\tilde h^\ell}\right)\right]\text ds \right),
	\end{equation}
	where $m \in \{0,1,\dots,\infty\}$ is the total number of innovations disclosed after time $t$, following history $h$ and assuming that no innovation is disclosed within $[t_h,t)$, $h^\ell$ is the history reached after the $\ell\text{th}$ disclosure (at time $t_\ell$) for each $0 < \ell \le m$, $h_0 = h$, $t_0 = t$, and $t_{m + 1} = \infty$ if $m < \infty$.
	$v^i_{\xi,h}(t)$ is agent $i$'s continuation payoff at time $t$ under $\xi$, following history $h$ and assuming that no innovation is disclosed within $[t_h,t)$.
	I Identify any strategy $\sigma^i$ of the game with forced disclosure with the strategy $(\sigma^i,\chi^*)$ where $\chi^*$ is the `full-disclosure' policy (that is, $\chi^*_h(t,z) = 1$ for all public histories $h$, $t > t_h$ and $z \ge 0$).
	Given a public history $h$ featuring $m \ge 0$ innovations, write $h \circ (t,z,i)$ for the public history that features $m+1$ innovations and extends $h$, and in which the last innovation is disclosed by agent $i$ at time $t$ and has size $z$.

	\begin{proposition}
		\label{proposition:ppe:dp}
		A strategy $\xi = (\sigma,\chi)$ of the game with disposal is a SSE if and only if, for any agent $i$, public history $h$, a.e.\ $t > t_h$, and $F$-a.e. $z$,
		\begin{align*}
			\sigma_h(t) &\in \argmax_{a \in [0,1]} a \lambda \mathbb{E}_F\left[\chi^i(t,\tilde z)\left(v^i_\xi(h \circ (t,\tilde z,i))-v^i_{\xi,h}(t)\right)\right] - c(a,x_h)
			\\\chi_h(t,z) &\in \argmax_{d \in \{0,1\}} d \left(v^i_\xi(h \odot (t,z,i)) - v^i_{\xi,h}(t) \right).
		\end{align*}	
		A strategy profile $\sigma = (\sigma^i)_{i = 1}^n$ of the game with forced disclosure is a PPE if and only if, for any agent $i$, public history $h$, and a.e.\ $t > t_h$,
		\begin{equation}
			\label{eq:ppe}
			\sigma^i_h(t) \in \argmax_{a \in [0,1]} a \lambda \left\{\mathbb{E}_F\left[v^i_\sigma(h \circ (t,\tilde z,i))\right]-v^i_{\sigma,h}(t)\right\} - c(a,x_h).
		\end{equation}
	\end{proposition}
	\Cref{proposition:ppe:dp} is standard; it follows easily from, e.g., Theorems 32.2 and 46.18 of \textcite{davis2018}.

	Recall \Cref{proposition:equilibrium:general} from \Cref{proof:equilibrium}.
	\begin{proposition}
		\label{proposition:sse_general}
		A strategy $\sigma$ of the game with forced disclosure is a SSE if and only if $\sigma = \alpha_e$.
		A strategy $\xi = (\sigma,\chi)$ of the game with disposal is a SSE if and only if $\sigma = \alpha_d$ and $\chi_h(t,z) = 1 \mathrel{(0)}$ for a.e.\ $t$ and $F$-a.e.\ $z$ such that $v_d(x+z) > \mathrel{(<)} v_d(x)$.
	\end{proposition}
	\begin{proof}[Proof of \Cref{proposition:sse_general}]
		Both `if' parts follow from \Cref{proposition:equilibrium:general}.
		For the `only if' parts, let $\Xi_e$ be the set of SSE of the game with forced disclosure and $\Xi_d$ be the set of SSE of the game with disposal.
		I will show that (i) $v^i_{\sigma,h}(t) =v_e(x_h)$ for any $\sigma \in \Xi_e$, history $h$ and $t > t_h$, and (ii) $v^i_{\xi,h}(t) = v_d(x_h)$ for any $\xi \in \Xi_d$, public history $h$ and $t > t_h$.
		It is easily seen that (i) yields the first `only if' part of \Cref{proposition:sse_general} and that (ii) yields the second, by \Cref{proposition:ppe:dp}.
		I prove (i) and (ii) simultaneously. 
		
		Fix $k \in \{d,e\}$, define $\bar v_k, \underline v_k : \mathbb R_+ \rightarrow \mathbb R$ by $\bar v_k(x) = \sup_{\xi \in \Xi_k} v^1_\xi(x)$ and $\underline v_k(x) = \inf_{\xi \in \Xi_k} v^1_\xi(x)$, and note that $\underline v_k(x_h) \le v^i_{\xi,h} \le \bar v_k(x_h)$ for all $\xi \in \Xi_k$, $h$ and $i$.
		Then, it suffices to show that $\bar v_k = v_k = \underline v_k$. 
		I only show that $\bar v_k = v_k$, as a similar argument yields $v_k = \underline v_k$.

		Note that $\bar v_k$ is upper-semicontinuous.
	  Since $b \le v_k \le \bar v_k \le v_*$ and $v_* = b$ on $[x_*,\infty)$ (\Cref{proposition:benchmark}), it is then enough to show that
		\begin{equation}
			\label{eq:strongly_symmetric_unique}
			\bar v_k(x)- v_k(x) \le L_k \bar v_k(x) - L_k v_k(x)
		\end{equation}
		for all $x$, where $L_k$ was defined in \eqref{eq:L}.%
		\footnote{It is clear that \eqref{eq:strongly_symmetric_unique} is sufficient if $k = e$; if $k = d$, the argument in footnote \ref{footnote:ranked} delivers sufficiency since, given any $\epsilon > 0$, there exists $\delta > 0$ such that $\bar v_k(x) \le b(x) + \epsilon$ for all $x \ge 0$ satisfying $\Pr_F(\bar v_k(x+\tilde z) > \bar v_k(x)) < \delta$.}
		Moreover, to obtain \eqref{eq:strongly_symmetric_unique}, it suffices to show that $v^i_\xi(x)- v_k(x) \le L_k \bar v_k(x) - L_k v_k(x)$ for all $\xi \in \Xi_k$, $i$ and $x$.

		Fix $\xi = (\sigma,\chi)$, $x$, and $i$, and assume without loss that $v^i_\xi(x) > v_k(x)$.
		Let	$t = \sup(\{0\} \cup \{s > 0: \sigma_x(r) \le \alpha_k(x) \text{ for a.e.\ } r \in (0,s)\})$ and set
		\begin{equation*}
			\phi = 
			\begin{cases}
				v^i_{\xi,x}(t) &\text{if $t < \infty$}\\
				0 & \text{otherwise}
			\end{cases}
			\qquad 
			\phi_* = 
			\begin{cases}
				L_k \bar v_k(x) - L_kv_k(x) + v_k(x) &\text{if $t < \infty$}\\
				0 & \text{otherwise.}
			\end{cases}
		\end{equation*}
		
		I first show that $\phi_* \ge \phi$.
		To this end, assume without loss that $t < \infty$, and note that $\alpha_k(x) < 1$.
		Note also that, for a.e.\ $s \ge t$ such that $\sigma_x(s) > \alpha_k(x)$,
		\begin{align*}
			L_k\bar v_k(x) - v^i_{\xi,x}(s) 
			&\ge \mathbb E_F\left[\chi_x(s,\tilde z)\left(v^i_\xi(x \circ (s,\tilde z,i))-v^i_{\xi,x}(s)\right) \right]
			\\&\ge c_1(\sigma_x(s),x)/\lambda 
			\ge c_1(\alpha_k(x),x)/\lambda 
			\ge L_kv_k(x) - v_k(x),
		\end{align*}
		where the first inequality holds since $\chi_x = 1$ if $k = e$, $\bar v_k(x+z) \ge v^i_\xi(x \circ (s,z,i))$ for all $z$, and $\bar v_k(x) \ge v^i_{\xi,x}(s)$;
		the second inequality follows from \Cref{proposition:ppe:dp} since $\sigma_x(s) > 0$; 
		the third holds since $\sigma_x(s) \ge \alpha_k(x)$; and the last follows from \eqref{eq:equilibrium} and \eqref{eq:disposal:equilibrium}, since $\alpha_k(x) < 1$.
		Choosing a sequence of $s$ with limit $t$ yields $\phi_* \ge v^i_{\xi,x}(t) = \phi$.

		Let $A$ be the set of all Lebesgue measurable $a: \mathbb{R}_+ \rightarrow [0,1]$. Given $a = (a_j)_{j = 1}^n \in A^n$, let $N(\cdot,a)$ be the CDF of the random time $\tau$ of the first innovation assuming that agents exert effort according to $a$ (with $\tau = \infty$ if no innovation is produced).%
		\footnote{That is, $N(\tau,a) = 1 - e^{-\lambda \int_0^\tau \sum_{j = 1}^n a_j}$ for all $\tau \in \mathbb{R}_+$.}
		For all (suitably integrable) $v : [0,t) \to \mathbb{R}$ and $w \in \mathbb{R}$, let
		\begin{equation*}
			\Phi(a,v,w) = \int \int_0^{s \wedge t} e^{-r}[b(x)-c(a_i(r),x)] \text dr + e^{-s \wedge t} (\mathbf{1}_{s < t} v(s) + \mathbf{1}_{s \ge t} w) N(\text ds,a),
		\end{equation*}
		be agent $i$'s ex-ante payoff at initial stock $x$, provided that (i) prior to the first innovation, agents implement $a$ on $[0,t)$, (ii) agent $i$'s continuation payoff at time $s < t$ is $v(s)$ if the first innovation occurs at time $s$, and (iii) agent $i$'s continuation payoff at time $t$ is $w$ if no innovation arises within $[0,t)$.
		I identify $a \in [0,1]$ with the constant map in $A$ taking value $a$ and $\hat a \in A$ with the element of $A^n$ whose entries are all equal to $\hat a$.
		Given $a,\hat a \in A$, let $a \oplus \hat a$ be the profile $(a_j)_{j = 1}^n$ such that $a_i = a$ and $a_j = \hat a$ for all $j \ne i$. 
		Define 
		\begin{equation}
			\label{eq:v_hat}
			\hat v = \max_{a_i \in A} \Phi\left(a_i \oplus \alpha_k(x),L_k \bar v_k(x),\phi_*\right)
		\end{equation} 
		and note that, if $k = d$, then $\max\left\{v^i_{\xi,x}(s),v^i_\xi(x \circ (s,z,j))\right\}$ is agent $i$'s continuation payoff at time $s \ge 0$ when the SSE $\xi$ is played, provided the initial stock is $x$ and the first innovation is produced by agent $j$ at time $s$, and has size $z$.
		Then,
		\begin{align*}
			v^i_\xi(x) 
			&= 
			\begin{cases}
				\Phi\left(\sigma_x,\textstyle \frac 1n \sum_{j=1}^n \mathbb E_F\left[v^i_\xi(x \circ (\cdot,\tilde z,j))\right],\phi\right)
				& \text{if $k = e$}\\
				\Phi\left(\sigma_x,\textstyle \frac 1n \sum_{j=1}^n \mathbb E_F\left[\max\left\{v^i_{\xi,x}(\cdot),v^i_\xi(x \circ (\cdot,\tilde z,j))\right\}\right],\phi\right) & \text{if $k = d$}
			\end{cases}
			\\&\le \Phi\left(\sigma_x,L_k \bar v_k(x),\phi_*\right) \le \hat v,
		\end{align*}
		where the first inequality holds since $v^i_{\xi,x}(s) \le \bar v_k(x)$ and $v^i_\xi(x \circ (s,z,i)) \le \bar v_k(x + z)$ for all $s$, $z$ and $j$, and $\phi \le \phi_*$; 
		and the second inequality holds since $\sigma_x \le \alpha_k(x)$ a.e.\ on $[0,t)$, $L_k \bar v_k(x) \ge b(x)$, and $L_k \bar v_k(x) \ge \phi_*$ (as $L_k v_k \ge v_k$, by \Cref{proposition:equilibrium:general}).%
		\footnote{In plain terms, the last inequality holds since $i$'s opponents exert higher effort if they implement $\alpha_k(x)$ instead of $\sigma_x$, and the first innovation before $t$ increases continuation payoffs.}

		Therefore, it suffices to show that $\hat v - v_k(x) \le L_k\bar v_k(x) - L_kv_k(x)$.
		To this end, let $\hat a_i \in A$ achieve the maximum in \eqref{eq:v_hat} and note that 
		\begin{align*}
			v_k(x) 
			&= \max_{a_i \in A} \Phi\left(a_i \oplus \alpha_k(x),L_kv_k(x),v_k(x)\right)
			\ge 
			\Phi\left(\hat a_i \oplus \alpha_k(x),L_kv_k(x),v_k(x)\right)
			\\\hat v - v_k(x) &\le [L_k \bar v_k(x)-L_kv_k(x)] \textstyle \int e^{-s \wedge t} N\left(\text ds,\hat a_i \oplus \alpha_k(x)\right) \le 
			L_k \bar v_k(x)-L_kv_k(x),
		\end{align*}
		where the second line follows from the first. 
	\end{proof}
	\begin{proof}[Proof of \Cref{proposition:ppe}]
		Fix a PPE $\sigma = (\sigma^i)_{i = 1}^n$.
		I will show that, for any history $h$, $\int_{t_h}^\infty \sum_{i = 1}^n \sigma^i_h(t)\mathrm{d}t > 0$ if $x_h < x_e$ and only if $x_h \le x_e$.
		Note first that 
		\begin{equation}
			\label{eq:ppe_cutoff}
			\lambda \{\mathbb{E}_F[b(x+\tilde z) - b(x)\} > \mathrel{(<)} c_1(0,x) \quad \text{for all $x < \mathrel{(>)} x_e$,}
		\end{equation}
		by \Cref{theorem:equilibrium}, since $b$ is concave and $c_1(0,x)$ is strictly increasing in $x$.

	 	For the `if' part of the aforementioned claim, suppose by means of contradiction that there exists a history $h$ such that $x_h < x_e$ and $\int_{t_h}^\infty \sum_{i = 1}^n \sigma^i_h = 0$.
		Note that, for any $i$ and any $t > t_h$, 
		\begin{equation*}
			\lambda\left\{\mathbb{E}_F\left[v^i_\sigma(h \circ (t,\tilde z,i))\right] - v^i_{\sigma,h}(t)\right\} \ge \lambda\{\mathbb{E}_F[b(x_h+\tilde z)]-b(x_h)\} > c_1(0,x_h),
		\end{equation*}
		contradicting \Cref{proposition:ppe:dp}, 
		where the first inequality holds since $v^i_\sigma(h \circ (t, z,i)) \ge b(x_h +  z)$ (as $\sigma$ is a PPE) and $v^i_{\sigma,h}(t) = b(x_h)$ (as $\int_{t_h}^\infty \sum_{i = 1}^n \sigma^i_h = 0$), and the second follows from \eqref{eq:ppe_cutoff} since $x_h < x_e$.

		For the `only if' part of the claim, I first show that $\int_{t_h}^\infty \sum_{i = 1}^n \sigma^i_h = 0$ for any history $h$ such that $x_h \ge x_*$.
		To this end, fix a history $h$ such that $x_h \ge x_*$ and note that $v^i_\sigma(h) = b(x_h)$ for each $i$, for otherwise $\sum_{i = 1}^n v^i_\sigma(h) > n b(x_h) = n v_*(x_h)$, where the inequality holds since $v^i_\sigma(h) \ge b(x_h)$ for each $i$ and the equality follows from \Cref{proposition:benchmark}.
		Then, for any $i$ and any $t > t_h$,
		\begin{equation*}
			\lambda\left\{\mathbb E_F\left[v^i_\sigma(h \circ (t,\tilde z,i))\right] - v^i_{\sigma,h}(t)\right\} 
			= \lambda\{\mathbb E_F[b(x_h+\tilde z)] - b(x_h)\} 
			< c_1(0,x_h)
		\end{equation*}
		where the inequality follows from \eqref{eq:ppe_cutoff} since $x_e < x_*$ (by \Cref{theorem:equilibrium}).
		Hence, $\int_{t_h}^\infty \sum_{i = 1}^n \sigma^i_h = 0$ by \Cref{proposition:ppe:dp}.

		Let $\hat x$ be the supremum of all $x \ge x_0$ such that $\sigma^i_h > 0$ for some $i$ and $h$ with $x_h \ge x$, assuming without loss that some such $x$ exists.
		It suffices to show that $\hat x \le x_e$.
		By the previous paragraph, $\hat x \le x_* < \infty$.
		Fix $\epsilon > 0$, a history $h$ such that $\hat x - \epsilon \le x_h \le \hat x$, and $i$ such that $\int \sigma^i_h > 0$.
		Then, there exists $t > t_h$ such that $\sigma^i_h(t) > 0$ and \eqref{eq:ppe} holds, so that 
		\begin{align*}
			c_1(0,x_h) \le c_1\left(\sigma^i_h(t),x_h\right) 
			&\le \lambda\left\{\mathbb E_F\left[v^i_\sigma(h \circ (t,\tilde z,i))\right] - v^i_{\sigma,h}(t)\right\}
			\\ &\le \lambda\left\{ \mathbb E_F\left[b(x_h+\tilde z) + \mathbf 1_{\tilde z \le \epsilon} \lambda n b(\mu)\right] - b(x_h)\right\},
		\end{align*}
		where the second inequality follows from \eqref{eq:ppe} since $\sigma^i_h(t) > 0$, and the third inequality holds since $v^i_{\sigma,h}(t) \ge b(x_h)$, $v^i_\sigma(h \circ (t,z,i)) \le b(x_h+z) + \lambda n b(\mu)$ for all $z$ (since $b$ is positive and concave), and $v^i_\sigma(h \circ (t,z,i)) = b(x_h+z)$ for all $z > \hat x - x_h$.
		Since $F$ is continuous, letting $\epsilon$ tend to $0$ yields that $\hat x \le x_e$, by \eqref{eq:ppe_cutoff}.
	\end{proof}
	\section{Proofs of \Cref{theorem:disposal:equilibrium,proposition:disposal_welfare}}
	\label{proof:disposal}
	I prove \Cref{theorem:disposal:equilibrium,proposition:disposal_welfare}. The proof of \Cref{theorem:disposal:equilibrium} shows that $\alpha_d$ has the properties claimed after the theorem. 

	Recall \Cref{proposition:equilibrium:general} from \Cref{proof:equilibrium}, \Cref{lemma:equilibrium:ge} from \Cref{proof:linear} and \Cref{proposition:sse_general} from \Cref{proof:ppe}.
	\begin{proof}[Proof of \Cref{theorem:disposal:equilibrium}]
		\Cref{proposition:equilibrium:general} yields a symmetric MPE $(\alpha_d,\delta_d)$ inducing continuation payoff $v_d$, such that $\alpha_d$ has the properties claimed after \Cref{theorem:disposal:equilibrium}, and $\delta_d(x,z) = 1$ if and only if $v_d(x+z) \ge v_d(x)$.

		To prove that $v_d \ge v_e$ and $\alpha_d \ge \alpha_e$, note that $L_dv_e(x) \ge L_ev_e(x) \ge b(x) \ge 0$ for all $x$, where the second inequality is strict if $x = 0$ and the third one if $x > 0$, since $v_e \ge b$ and $b$ is increasing with $b'(0) > 0$.
		Then 
		\begin{equation}
			\label{eq:disposal_ge}
			v_d(x) \ge \Gamma(p(x,L_ev_e(x)),x,L_ev_e(x)) = v_e(x)
		\end{equation}
		and $\alpha_d(x) \ge p(x,L_ev_e(x)) = \alpha_e(x)$ for all $x$, where the inequalities follow from \Cref{lemma:equilibrium:ge} and the equalities from \Cref{proposition:equilibrium:general}.

		The fact that any SSE $\xi = (\sigma,\chi)$ of the game with disposal satisfies $\sigma = \alpha_d$ and $v^i_\xi(h) = v_d(x_h)$ for all $i$ and $h$ follows from \Cref{proposition:sse}.

		Fix $\hat x > 0$ and suppose that $v_e$ is increasing on $[\hat x,\infty)$.
		To show that $\delta_d(x,z) = 1$ and $\alpha_d(x)=\alpha_e(x)$ for all $x \ge \hat x$ and $z \ge 0$, note that the symmetric strategy profile induced by $\alpha_e$ and the full-disclosure policy is a SSE of the game with disposal having initial stock $x_0 = \hat x$, so that $\alpha_e = \alpha_d$ and $v_d = v_e$ on $[\hat x,\infty)$.
		Hence, $v_d$ is increasing on $[\hat x,\infty)$, so that $\delta_d(x,z) = 1$ for all $x \ge \hat x$ and $z \ge 0$.

		Suppose now that $v_e$ is not increasing on $[\hat x,\infty)$.
		It remains to show that $v_d(\hat x) > v_e(\hat x)$ and that $v_d$ is not increasing on $[\hat x,\infty)$.
		For the former, note that $\hat x < x_e$ and $v_e$ is not increasing on $[\hat x,x_e]$, since $v_e = b$ on $[x_e,\infty)$ by \Cref{theorem:equilibrium}.
		Because $F$ has convex support including $0$, it follows that $\hat x \in X_m$ for some $m \in \{0,1,\dots\}$, where $X_0 = \{y \in [0,x_e):\Pr_F(v_e(y+\tilde z) < v_e(y)) > 0\}$ and $X_m = \{y \in [0,x_e) : \Pr_F(y + \tilde z \in X_{m-1}) > 0\}$ for all $m \in \mathbb{N}$.
		Then, it suffices to show that $v_d > v_e$ on $X_m$ for all $m$.
		Fix $m$ and $x \in X_m$ and note that $L_d v_d(x) \ge L_e v_d(x) > L_e v_e(x)$, where the first inequality holds since $v_d \ge v_e$, and the second since we may assume without loss that $v_d > v_e$ on $X_{m-1}$ if $m > 0$.
		Note also that $p(x,L_ev_e(x)) = \alpha_e(x) > 0$, where the inequality follows from \Cref{theorem:equilibrium} since $x < x_e$.
		Then, the inequality in \eqref{eq:disposal_ge} is strict, by \Cref{lemma:equilibrium:ge}. 

		To prove that $v_d$ is not increasing on $[\hat x,\infty)$, note that if $v_d$ were increasing on $[\hat x,\infty)$ then $\alpha_d$ would induce a SSE of the game with forced disclosure having initial stock $x_0 = \hat x$, and this would contradict \Cref{proposition:sse} since $v_d(\hat x) > v_e(\hat x)$.
	\end{proof}
	\begin{proof}[Proof of \Cref{proposition:disposal_welfare}]
		Suppose that \eqref{eq:disposal_welfare} holds.
		Let $\beta = \mathbb{E}_F[\max\{b(x+\tilde z) - \mathbb{E}_{G^x}[b(\tilde y)],0\}]$ and $\bar u = b(x)-c(1,x)$, and fix $\hat a > 0$ such that $\lambda\beta > c_1(\hat a,x)$.
		It suffices to show that $v_d(x) > \mathbb{E}_{G^x}[b(\tilde y)]$ if $n > (\mathbb{E}_{G^x}[b(\tilde y)]-\bar u)/(\lambda \hat a \beta)$, since the last part of \Cref{proposition:disposal_welfare} then follows by definition of $G^x$.
		To this end, note that $v_d(x) > \mathbb{E}_{G^x}[b(\tilde y)]$ if $\alpha_e(x) < \hat a$, since 
		\begin{equation*}
			\mathbb{E}_F[\max\{b(x+\tilde z)-v_d(x),0\}] \le \mathbb{E}_F[\max\{v_d(x+\tilde z)-v_d(x),0\}] \le c_1(\hat a,x)/\lambda < \beta,
		\end{equation*}
		where the second inequality follows from \eqref{eq:disposal:equilibrium}. 
		If $\alpha_e(x) \ge \hat a$, then $n \le (\mathbb{E}_{G^x}[b(\tilde y)]-\bar u)/(\lambda \hat a \beta)$ provided that $v_d(x) \le \mathbb{E}_{G^x}[b(\tilde y)]$ since, in this case,  
		\begin{align*}
			\mathbb{E}_{G^x}[b(\tilde y)] \ge v_d(x) &= b(x)-c(\alpha_e(x),x) 
			\\& \quad + \lambda n \alpha_e(x) \mathbb{E}_F[\max\{v_d(x+\tilde z)-v_d(x),0\}]   
			\ge \bar u + \lambda n \hat a \beta, 
		\end{align*}
		where the equality follows from \eqref{eq:disposal:equilibrium}.
	\end{proof}
	\end{appendices}
	\end{document}